%% file: NSBP06.tex
\documentclass[12pt,epsf]{article}
\usepackage{epsfig}

\newcommand{\onefigure}[2]{\begin{figure}[htbp]
\begin{center}\leavevmode\epsfbox{#1.eps}\end{center}\caption{#2\label{#1}}
\end{figure}}

\renewcommand{\thanks}[1]{\footnote{#1}} 

\newcommand{\be}{\begin{equation}}
\newcommand{\ee}{\end{equation}}
\newcommand{\bea}{\begin{eqnarray}}
\newcommand{\eea}{\end{eqnarray}}

\begin{document}

\pagestyle{empty}

\bigskip\bigskip
\begin{center}
{\bf \large Evidence for a Cosmological Phase Transition
From the Dark Energy Scale\footnote{\baselineskip=12pt Talk presented at the 2006 Annual
Conference of the National Society of Black Physicists and National Society of Hispanic
Physicists, Feb. 15-18, San Jose, California.}}
\end{center}

\begin{center}
James Lindesay\footnote{e-mail address, jlslac@slac.stanford.edu} \\
Computational Physics Laboratory \\
Howard University,
Washington, D.C. 20059 
\end{center}

\begin{center}
{\bf Abstract}
\end{center}
\input{abstNSBP.tex}
\newpage

\setcounter{equation}{0}
\input{intro06.tex}

\setcounter{equation}{0}
\input{Decoh06.tex}

\setcounter{equation}{0}
\input{subfluc.tex}

\setcounter{equation}{0}
\input{scenar06.tex}

\setcounter{equation}{0}
\input{rhopt2.tex}

\setcounter{equation}{0}
\input{discuss2.tex}

\begin{center} {\bf Acknowledgment}
\end{center}

The author gratefully acknowledges useful
discussions with the following:
Stephon Alexander, James Bjorken, Stan Brodsky, E.D. Jones, L.H. Kauffman, T.W.B. 
Kibble, Walter Lamb, H. Pierre Noyes, Michael Peskin, and Lenny Susskind.

\input{bib3.tex}
\end{document}

%% file: abstNSBP.tex
A finite vacuum energy density implies the existence of a
UV scale for gravitational modes.  This gives a phenomenological
scale to the dynamical equations governing the cosmological
expansion that must satisfy constraints consistent with quantum
measurability and spatial flatness.
Examination of these constraints for the observed dark energy density
establishes a time interval from the transition to the present,
suggesting major modifications from the thermal equations
of state far from Planck density scales.
The assumption that a phase transition initiates the radiation
dominated epoch
is shown under several scenarios to produce fluctuations to
the CMB of the order observed.
Quantum measurability constraints (eg. uncertainly relations)
define cosmological scales
bounded by luminal expansion rates.
It is shown that the dark energy can consistently be interpreted
as being due to the vacuum energy of collective gravitational modes
which manifest as the zero-point motions of coherent Planck
scale mass units prior to the UV scale onset of gravitational quantum de-coherence
for the cosmology.

%% file: intro06.tex
\section{Introduction \label{intro}}
\indent

Recent evidence for the acceleration of the universe
suggests that the dominant component of today's
cosmology is in the form of dark energy.  Many
feel that understanding the nature of this dark energy ranks
among the most compelling of all outstanding problems
in physical science.  This presentation will suggest
conceptual frameworks for investigating this intriguing
problem.

For this treatment, it is assumed that the 
behaviors of the known particles do not qualitatively change
up to energies in the TeV range.  It is
further assumed that the current understanding of general
relativity as a gravitational theory with
thermal energy content is adequate over the same
range, and consequently that the cosmological Friedman-Lemaitre
(FL) dynamical equations are reliable guides once 
the observational regime has been reached where the homogeneity and
isotropy assumptions on which those equations are based become
consistent with astronomical data to requisite accuracy. 
The elementary particle behaviors are assumed to apply
locally on coordinate backgrounds
with cosmological curvature. There is direct experimental
evidence that quantum mechanics does apply in the background space
of the Schwarzschild metric of the Earth from experiments by Overhauser and
collaborators\cite{Over74,Over75}. These experiments show that the coherent
self-interference pattern of single neutrons changes as expected
when the plane of the two interfering paths is rotated from being
parallel to being perpendicular to the ``gravitational field" of
the Earth.  These experiments provide a verification of the principle of equivalence
for quantum systems (at least in stationary geometries). 

It is expected that during some period in the past, quantum coherence of
gravitating systems should have qualitatively altered the dynamics
of the cosmology.  Often, the onset of the importance of quantum
effects in gravitation is taken to be at the Planck scale.
However, as is the case with Fermi degenerate stars, this need not
be true of the cosmology as a whole.  For the same reason that
neutron star densities are not necessary for quantum coherence
effects to manifest in superfluid helium, Planck scale densities need not
be necessary for macroscopic coherence effects to manifest for Planck
scale masses.  Quantum coherence
refers to the entangled nature of quantum states for space-like separations.
This is made evident by superluminal correlations (without the
exchange of signals) in the observable behavior of such quantum states.
Note that the exhibition of quantum coherent behavior for gravitating systems does
\emph{not} require the quantization of the gravitation field.

The equations which govern the (spatially flat) cosmological expansion satisfy spatial
scale invariance, but not temporal scale invariance, due to the behavior
of the intensive energy densities which drive the dynamics.  However, a
specific scale for gravitational modes can be defined by the onset of
a microscopic critical density defining a phase transition.  This 
density couples the macroscopic cosmological dynamics to the
microphysical  phenomena associated with particle dynamics.
Collective gravitational modes which satisfy quantum measurability
constraints of higher energies than that defined by the phase transition
microscopically thermalize once this threshold is reached, while
super-horizon modes satisfy those measurability constraints at later times.
The consequences of the existence of this coherent gravitational UV
scale is what will be explored in this presentation.

As mentioned,
the Friedmann-Lemaitre (FL) equations for an ideal fluid governing cosmological
evolution is spatially scale invariant unless there is spatial curvature,  
with the time scale
determine by intensive energy densities.  The oldest galactic clusters are of the
order $12Gyrs$, so that any estimations of the age of the expansion
must be greater than this.  Using current
observations, the period of present expansion has been estimated to
have a duration of about $13.7 \pm 0.2 Gyrs$.  The cosmological critical density (defined
in terms of the present Hubble expansion rate as the maximum energy density
which would not close the universe, resulting in eventual collapse) is
of the order $\rho_{crit} \approx 0.9 \times 10^{-29} g \: c^2/cm^2 \approx
0.5 \times 10^{-5}GeV/cm^3$.  The relative densities of the various constituent
components are of the order $\Omega_\gamma \sim 4.9 \times 10^{-5}$,
$\Omega_{baryons} \sim 0.04$, $\Omega_{dark \, matter} \sim 0.22$, and
$\Omega_{dark \, energy}\sim 0.73$.  
The radiation component corresponds to about 413 photons/$cm^3$
with a cosmological microscopic
entropy density dominated by these photons $s_\gamma \approx
2905 k_B /cm^3$.  The universe is evidently very hot with regards to the baryons,
with an entropy per baryon of the order $10^{10} k_B$ for the universe
(compared with $10^{-2} k_B$ for the sun and of the order $k_B$ for neutron stars).
This large entropy per baryon is related to the baryon-antibaryon asymmetry
which ultimately results in the small component of dust which planets
and stars are made of.

The observed cosmic microwave background radiation
is extremely uniform.  When variations due to motions of
our galaxy and known sources are subtracted, the primordial
variations are seen to be of the order $10^{-5}$ (Figure \ref{CMBfluc}). 
\onefigure{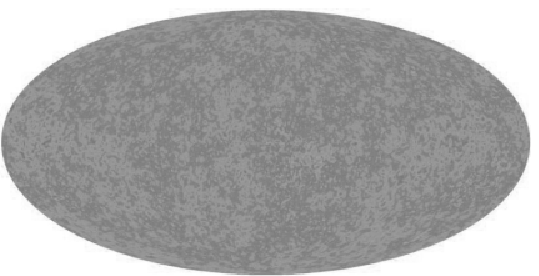}{CMB radiation contrast 30000}
This early remnant from the big bang is responsible for a few percent
of the thermal noise appearing as ``snow" in broadcast television.
The surface of last scattering imaged by the Cosmic Microwave Background (CMB)
radiation is a snapshot of the universe during the period of the formation of
hydrogen atoms from the prior hot plasma of the constituent particles. 
Because a hot hydrogen plasma is essentially opaque to
electromagnetic radiation, this region of last scattering acts as a
fog to our (electromagnetic) view of the universe at earlier times.
It was formed when the universe was about 300K years old, at a black body
temperature $k_B T_{LS}$ corresponding to about 0.3 eV (about $3000 ^o K$), at
a time when cosmological scales were about $z_{LS} \sim 1100$ times smaller
than today.  The temperature has been suppressed by this expansion factor,
resulting in the relatively cold, dark night sky observed today, which glows as
a hot, hydrogen plasma cooled to the
present $2.74 ^o K$ microwave background.
The fluctuations from uniformity in the CMB map of the sky are of the
order ${\delta \rho_{CMB} \over \rho_{LS}} \sim 10^{-5}$.  For comparison,
typical fluctuation scales observed today are given by
${\delta \rho_{stars} \over \rho_o} \sim 10^{30}$ for stars,
${\delta \rho_{galaxies} \over \rho_o} \sim 10^{5}$ for galaxies,
${\delta \rho_{clusters} \over \rho_o} \sim 10^{2}$ for clusters, and
${\delta \rho_{superclusters} \over \rho_o} \sim 1$ for superclusters. 
The spectrum of the CMB radiation and power spectrum of the fluctuations
are represented in Fig. \ref{CMBbbr}.
\onefigure{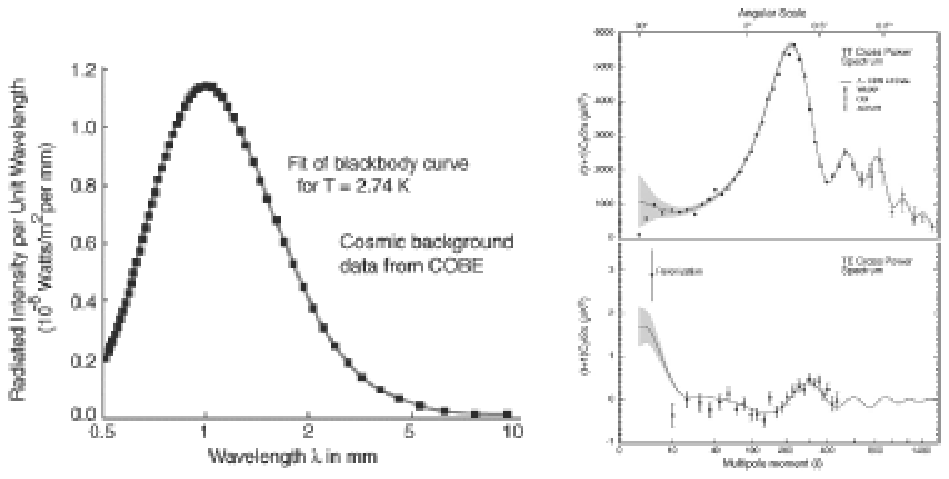}{Figure on left demonstrates extreme agreement of 
observed CMB intensity spectrum with blackbody curve.   Figure on right
demonstrates observed fluctuation power spectrum vs angular mode.}

The small fluctuations are believed to have propagated as 
primordial acoustic waves prior to the de-coupling of those photons from
ionized hydrogen.  In the early universe, after many Compton
scatterings, charges and photons would reach statistical
equilibrium, with the photons effectively having a non-vanishing
chemical potential since Compton scattering conserves
photon number.  The wave equation
results from combining photon number conservation 
$n_\gamma ' + \vec{\nabla} \cdot (n_\gamma \vec{v}_\gamma)=0$
to Euler's equation$(\rho_\gamma + P_\gamma) \vec{v}_\gamma '
=-c^2 \vec{\nabla}P_\gamma$,
where the temporal derivatives indicated by primes are with respect to
conformal time (which gives the proper distance
a photon will have traveled when multiplied by the dimensional
cosmological scale factor).  For a photon gas, the number density is
proportional to the cube of the temperature, while the energy density
is proportional to the fourth power of the temperature.  The equation
of state of a photon gas satisfies $P_\gamma = {1 \over 3}\rho_\gamma$.
Since the dominant component of the observed photon velocity will be
in the direction of the propagation vector $k$ of the modes, small
dimensionless temperature fluctuations $\Theta$ will satisfy
$3 \Theta_k '' + k^2 c^2 \Theta_k =0$, which indicates that the (conformal) speed
of sound in the acoustic wave was given by $c/\sqrt{3}$.

The (luminal) horizon problem
examines the large scale homogeneity and isotropy of the observed universe. 
Examining the
distance  photons can have traveled during the evolution of the universe prior to
decoupling from ionized matter,
which turns out to be about 1/100 the distance they have traveled
to present time,
the subsequent expansion should image light in the cosmic microwave background 
from $100^3=10^6$ luminally disconnected regions. 
Yet, uniformity of temperature and
angular correlations of the fluctuations across the whole sky have been
accurately measured by several experiments\cite{WMAP}. 
Even more intriguing,
the observation of these correlations provides evidence for
a space-like coherent phase associated with the
cosmological fluctuations that produced the CMB.
It is commonly assumed that
a period of inflation in the early universe is necessary to explain
these acausal phenomena.  However, a classical inflation cannot explain space-like
coherence of a macroscopic phase.  Indeed, only quantum phenomena exhibit
properties of space-like phase coherence of otherwise stochastic processes.

Using the usual vacuum state in Minkowski space-time, the equal time correlation function 
${<vac|\Psi(x,y,z,t) \, \Psi(x',y',z',t)|vac>}$ of a quantum field $\Psi(\vec{x})$
does not vanish for space-like separations.  For example, for massless scalar fields, 
$<vac|\Psi(\vec{x}) \, \Psi(\vec{y})+\Psi(\vec{y}) \, \Psi(\vec{x})|vac>={1 \over 4 \pi^2 s^2}$, 
with space-like phase coherence, where
the proper distance satisfies $s^2 = |\mathbf{x}-\mathbf{y}|^2 -
(x^0-y^0)^2$, which falls off with the inverse square of the distance between the points. 
Since the vacuum expectation value of the field $\Psi$ vanishes in the usual case, this
clearly requires space-like correlations, ie
\be
<vac|\Psi(\vec{x}) \, \Psi(\vec{y})+\Psi(\vec{y}) \, \Psi(\vec{x})|vac>\not= 
2 <vac|\Psi(\vec{x}) |vac><vac| \Psi(\vec{y})|vac> .
\ee
However, since the commutator of the field \emph{does} vanish for
space-like separations, a measurement at $\vec{y}$ cannot change
the probability distribution at $\vec{x}$.
In the approach here taken, global gravitational
coherence solves (or defers) the
horizon problem because the correlations of a macroscopic quantum
system \emph{are} space-like; it is hypothesized that the same
will be true of any type of quantum dark energy or phases of gravitational significance.
Quantum systems additionally satisfy
measurability constraints resulting in the usual uncertainty
relations. 
It will be argued that the equilibration of microscopic interactions
can only occur on cosmological scales consistent with quantum
measurement constraints\cite{SLACTalk}.

Einstein's equation is assumed to accurately describe the
evolution of the thermal universe.  The equation
relates a geometrically conserved combination of
curvature tensors to the dynamically conserved energy-momentum tensor:
\be
G_{\mu \nu} \equiv \mathbf{R}_{\mu \nu} -
{1 \over 2} g_{\mu \nu} \mathbf{R} =
{8 \pi G_N \over c^4} T_{\mu \nu} + \Lambda g_{\mu \nu}.
\label{EinsteinEqn}
\ee
The cosmological constant term, originally considered a
a blunder included to justify a stationary state universe,
has recently been resurrected in an effort to explain
an apparent acceleration of the expansion rate of
the universe.  For the present discussion, it will be
placed on the right hand side of Einstein's equation,
implying a closer connection to the microphysics
of the energy-momentum dynamics rather than
an inherent aspect of the geometry.

Type 1a supernovae are standard candles with
a 12\% dispersion in brightness and a known
temporal profile.  These supernovae form as the mass
from a binary companion of a white dwarf accretes
onto the white dwarf
until it reaches the Chandrasekhar limit associated
with gravitational saturation of the
Fermi degeneracy pressure of the electrons,
when it explodes in a standard way.  High redshift
type 1a supernovae are about 25\%  fainter than
would be expected in a decelerating universe.
If this faintness were due to macroscopic ``gray" dust, the finite
grain size would not look gray in the infrared
(similarly the sunset looks red).
No sign of such an effect has been observed.
Also, dust effects would be expected to modify
more distant galaxies even moreso than the
nearer galaxies, yet these galaxies do appear
to decelerate.
Thus, the luminosities of distant type Ia
supernovae show that the rate of expansion of the
universe has been accelerating for about 6 
giga-years\cite{TypeIa}. 
Figure \ref{darkE} demonstrates that nearer galaxies seem
to be accelerating their recession, whereas further galaxies
are decelerating\cite{Paddy1,Paddy2}.
\onefigure{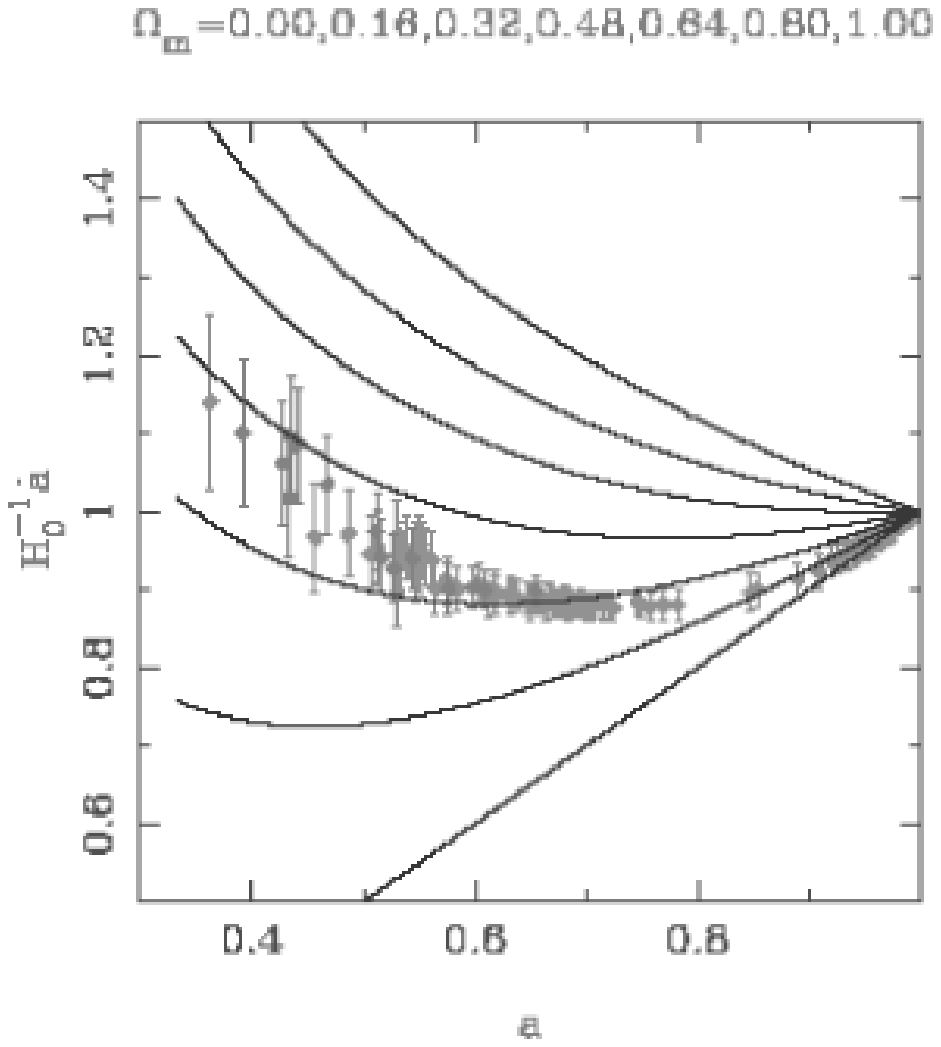}{Plot of observed scale expansion rate vs scale.  The
theoretical plots indicate expected behaviors for cosmologies
with $\Omega_m$=0, 0.16, 0.32, 0.48, 0.64, 0.80, and 1}
One implies from this that the scale was decelerating in the more
distant past, and is presently accelerating.
This conclusion is independently supported
by analysis of the Cosmic Microwave Background (CMB)
radiation\cite{PDG,WMAP}.  The CMB indicates a
spatially flat universe with the total density relative
to the cosmological critical density given by $\Omega_{total}\cong 1$.
When examining the features of the anisotropies in the CMB,
features in an open (closed) universe later appear to be
closer (further) than they actually were at the time observed.
However, the observed location of the first acoustic peak
appropriate to the expected sound velocity constrains the
space to be very nearly flat.  The height of the first acoustic
peak is also sensitive to the matter density.  Careful measurements
including totally unrelated processes (such as the deuterium/hydrogen
ratio) give relativistic radiation, baryonic, and total
pressureless mass (including dark matter) relative densities
adding up to about $0.27 < 1$.  This means that the missing relative energy
density must be of the order 0.73.  Thus
both the standard candle luminosity and CMB structure
results are independently in quantitative
agreement with a (positive) cosmological constant fit to the data.
The existence of a finite cosmological constant / dark energy density
defines a length scale
that should be consistent with those scales generated by the
microscopic physics, and must be incorporated in any
description of the evolution of the universe.   

The density of states for quantized modes
can be shown to be independent of the
shape of the boundary region\cite{densityofstates}
\be
\rho_\Lambda = {g_\epsilon \over b_\Lambda} {1 \over 2 \pi^2}
\int _{k_{IR}} ^ {k_{UV}} \hbar k v_p k^2 dk \cong 
{g_\epsilon \over b_\Lambda} {\hbar v_p \over 8 \pi^2} k_{UV} ^4
\label{darkenergy}
\ee
where $v_p$ is the phase velocity of the modes,
$b_\Lambda$ is a factor relating how much of the
vacuum energy makes up the dark energy ($b_\Lambda=2$ if
all vacuum energy is dark energy) and $v_p$
is the phase velocity of the modes.
 
Quantum measurement and flatness arguments will later be made for quantized
energy units E requiring that the scales associated with those
energies, $R_E$, must satisfy $\dot{R}_E \leq c$, which relates
$k_{UV}$ to the cosmological expansion.
The gravitationally coherent cosmological dark energy modes
begin to decouple from the thermalized
energy density in the Friedmann-Lemaitre(FL) equations when the
FRW UV scale expansion rate is no longer supra-luminal. 
A key assumption in this presentation is that this decoupling corresponds
to a phase transition from some form of a macroscopic coherent state
to local states understood in terms of late time observations.
If there is a UV scale associated with the gravitational modes,
the dark energy can be associated with the vacuum energy of
those modes.  

We will briefly discuss the nature of vacuum energy. 
Vacuum energy is \emph{not} due to the background
fluctuations of the basis of a particular perturbative expansion
(like vacuum polarization or mass renormalization).  The
physical manifestations must be independent of any basis of expansion
of the physical states.
The question then is, dark energy corresponds to the
vacuum energy of \emph{what}?  
The known particle spectrum would have far too large vacuum energy to correspond with observations. 
Rather than focusing on vacuum energy, we will examine the zero-point motions of quantum correlated 
gravitating sources.
The basic assumption is that the
quantum zero-point energies whose effects on the subsequent
cosmology are fixed by the phase transition
are to be identified with the cosmological constant, or ``dark energy".

Since we identify ``dark energy" as a
particular ``vacuum energy" driven by zero-point motions, it might be
illuminating to examine the physics behind other systems that manifest
vacuum energy.  One such physical system is the Casimir effect\cite{Casimir}.
Casimir considered the change in the vacuum energy due to the placement of
two parallel conducting plates separated by a distance $a$. 
He calculated an energy per unit area of the form
\be
{ {1 \over 2} \left ( \sum_{modes} \hbar c k_{interior} -
\sum_{modes} \hbar c k_{exterior}  \right ) \over A} \: = \:
- {\pi^2 \over 720} {\hbar c \over a^3}
\label{CasimirEnergy}
\ee
resulting in an attractive force given by
\be
{F \over A}  \: = \:
- {\pi^2 \over 240} {\hbar c \over a^4}
\: \cong \: {- 0.013 \, dynes \over (a / micron)^4  } cm^{-2},
\ee
independent of the charges of the sources.  Although
the effect does not depend on the electromagnetic
coupling strength of the sources, it does depend on
the nature of the interaction and configuration. 
Lifshitz and his collaborators\cite{Lifshitz} demonstrated that the Casimir force
can be thought of as the superposition of the van der Waals attractions
between individual molecules that make up the attracting media.  This allows
the Casimir effect to be interpreted in terms of the zero-point motions of the
sources as an alternative to vacuum energy of the associated quanta. 
At zero temperature, the coherent zero-point motions of
source currents on opposing plates correlate in a manner resulting
in a net attraction, whereas if the motions were independently random,
there would be no net attraction.  On dimensional grounds, one can
determine that the number of particles per unit area undergoing
zero-point motions that contribute to the Casimir result vary as $a^{-2}$.
Boyer\cite{Boyer} and others subsequently
derived a repulsive force for a spherical geometry of the form
\be
{1 \over 2} \left ( \sum_{modes} \hbar c k_{interior} -
\sum_{modes} \hbar c k_{exterior}  \right )  \: = \:
{ 0.92353 \hbar c \over a   }.
\ee
This shows that the change in electromagnetic vacuum energy \textit{is}
dependent upon the geometry of the boundary conditions,
although it does not depend on the detailed couplings of the
involved interactions.  Both
predictions have been confirmed experimentally.  It is important to note
that this energy grows inversely with the geometric scale
$E_{Casimir} \sim {\hbar c \over a}$.

Others have likewise noted that correlated zero-point motions can be used to describe
vacuum energy effects in the Casimir effect (see Fig \ref{vacZPM}).  As expressed by
Daniel Kleppner\cite{Kleppner},
\begin{quotation}
The van der Waals interaction is generally described in terms of
a correlation between the instantaneous dipoles of two atoms
or molecules.  However, it is evident that one can just as easily
portray it as the result of a change in vacuum energy due to
an alteration in the mode structure of the system.  The two
descriptions, though they appear to have nothing in common,
are both correct.
\end{quotation}
\onefigure{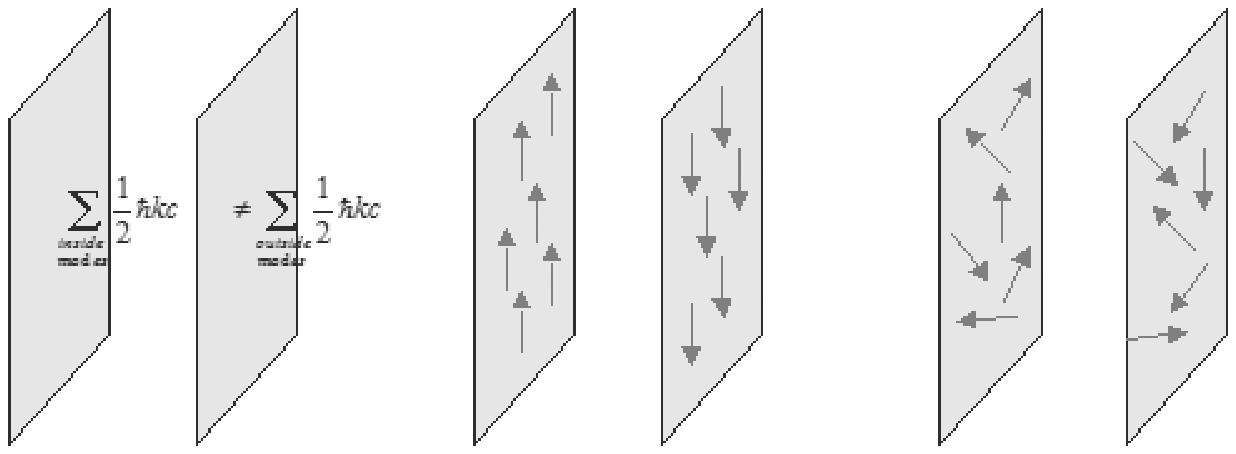}{The first diagram explains the Casimir pressure in terms
of a fewer number of modes (denumerably infinite) between the plates as
compared to the number of vacuum modes in the external region.  The
second diagram demonstrates an attraction due to space-like correlated
Van der Waals induced polarizations resulting in net attraction between
the sources.  Uncorrelated polarizations demonstrated in the third diagram
result in no net attraction.}
Additionally, as pointed out by Wheeler and Feynmann\cite{Wheeler},
and others\cite{LN2},
one cannot unambiguously separate the properties of
fields from the interaction of those fields with their sources
and sinks.
Since there are no manifest boundaries in the description of
early cosmology presented here, it is more convenient
to examine the effects of any gravitational ``vacuum energy'' in terms of
the correlated zero-point motions of the sources of those
gravitational fields.  In what follows, the zero-point motions of
coherent sources will be considered to correspond to the
vacuum energies of the associated quanta. 

Another system which manifests physically measurable effects due to zero-point energy is liquid $^4$He.
One sees that this is the case by noting that atomic radii are related to atomic volume $V_a$
(which can be measured) by $R_a \sim V_a ^{1/3}$.  The uncertainty relation gives momenta of the order
$\Delta p \sim \hbar / V_a^{1/3}$.  Since the system is non-relativistic, one estimates the zero-point kinetic
energy to be of the order $E_o \sim {(\Delta p)^2 \over 2 m_{He}} \sim {\hbar ^2 \over 2 m_{He} V_a ^{2/3}}$.
The minimum in the potential energy is located around $R_a$, and because of the low mass of $^4$He, the
value of the small attractive potential is comparable to the zero-point kinetic energy.  Therefore, this bosonic
system forms a low density liquid at densities much less than those associated with the nuclear
masses involved
(The lattice spacing for solid helium is expected to be even
smaller than the average spacing for the liquid,
which means that a large external pressure is necessary to
overcome the zero-point energy in order to form solid helium).

Applying this reasoning to relativistic gravitating mass units with quantum coherence within the
volume generated by a Compton wavelength $\lambda_m ^3$, the zero point momentum is expected
to be of order $p \sim {\hbar \over V^{1/3}} \sim {\hbar \over \lambda_m}$.  This gives a zero point
energy of order $E_0 \approx \sqrt{2} m c^2$.  If we estimate a mean field potential from the
Newtonian form $V \sim -{G_N m^2 \over \lambda_m}=-{m^2 \over M_P ^2}mc^2<<E_0$, it is evident that
the zero point energy would dominate the energy of such a system.

That vacuum energy can be thought of as resulting from
the zero-point motions of the sources is also supported by the calculation of Bohr
and Rosenfeld\cite{BohrRosenfeld}, who minimized the effect of a 
classical measurement of an electric field averaged over a finite volume
on the value of a magnetic field at right angles averaged over (i) a
non-overlapping volume, and (ii) an overlapping volume, and {\it vice versa}. 
When this minimum disturbance of sources is put equal to the minimum uncertainty
which the uncertainty principle allows for the measurement they
reproduce the result of averaging the quantum mechanical commutation
relations over the corresponding volumes.  Such arguments can be extended
to the corresponding case when the sources and detectors are
gravitational.

%% file: Decoh06.tex
\section{Cosmological Scale De-coherence From Dark Energy
\label{decohere2}}
\indent

The Friedman-Robertson-Walker (FRW) metric for a homogeneous
isotropic cosmology is given by
\be
ds^2 \: = \: c^2 dt^2 \, + \, R^2 (t) \left (
{dr^2 \over 1-\kappa \, r^2} + r^2 d \theta ^2
+ r^2 sin^2 \theta \, d \phi ^2
\right ) .
\label{FRWmetric}
\ee
Here the FRW scale factor $R$ is taken to have dimensions of length
(not to be the dimensionless scale relative to the present scale
$a(t) \equiv R(t)/R_o$), with $dr$ being dimensionless.

Our approach will be to start from well
understood macrophysics,
assume that the physics of a cosmological phase transition defines an
FRW scale parameter, and examine
cosmological physics at that time with regards to
the physical consistency of the thermal state
of the cosmology. For times after that transition
there is general confidence that well
understood micro- \emph{and} macro-physics are valid.
The FRW scale parameter must be expressed in terms of
scales relevant to microscopic physics.

\subsection{Quantum Measurability Constraints on Scale Expansion}
\indent

Substitution of the FRW metric
Eq. \ref{FRWmetric} into the Einstein Field equation \ref{EinsteinEqn} driven by
an ideal fluid result in the Friedmann-Lemaitre(FL) equations.
The FL equations,
which relate the rate and acceleration of the expansion to the 
fluid densities, are given by
\be
H^2 (t) \: = \: \left ( {\dot{R} \over R} \right ) ^2 \: = \: {8 \pi G_N \over 3 c^2}
\left ( \rho + \rho_\Lambda  \right ) \, - \, {\kappa c^2 \over R^2}
\label{Hubble_eqn}
\ee
\be
{\ddot{R} \over R} \: = \:
-{4 \pi G_N \over 3 c^2} (\rho + 3 P - 2 \rho_\Lambda),
\label{acceleration}
\ee
where $H(t)$ is the Hubble expansion rate, the dark energy density is given by
$\rho_\Lambda \: = \: {\Lambda c^4 \over 8 \pi G_N}$,
$\rho$ represents the FL fluid energy density,
and $P$ is the pressure.  These equations combine to give
the 1st Law of Thermodynamics for adiabatic expansion
$d(\rho R^3) = -P d(R^3)$.
The term which involves the spatial curvature $\kappa$ has
explicit scale dependence on the FRW parameter $R$.
The dark energy density makes a negligible contribution to
the FL expansion during early times, but becomes
significant as the FL energy density
decreases due to the expansion of the universe.

The scale evolution can be explicitly solved for relevant
cosmological energy content.  It is convenient to define
conformal time, which involves a coordinate transformation
insuring that light cones have a slope of unity:
\be
\begin{array}{c}
ds^2 \: = \: R^2 \left [ -d \eta^2
+ dr^2 + r^2 (d \theta^2 + sin^2 \theta d\phi^2)
\right ] \\ \\
d \eta = {c dt \over R(t) }
\end{array}
\ee
Light travels on null geodesics, making the conformal time
equivalent to the (dimensionless) distance traveled by
photons in a given time interval.

Driven by constant energy density,
an inflationary scale evolves according to
\be 
R(t) = R_I e^{H_I (t-t_I)} \quad , \quad
 {1 \over R(\eta)} = {1 \over R_I} 
-{H_I \over c} (\eta - \eta_I) .
\ee
The energy density of radiation varies as $R^{-4}$, so
a radiation dominated scale satisfies
\be
R^2(t) = R_{RD}^2 + 2 R_{RD} c (t-t_{RD}) 
\quad , \quad
 R(\eta) = R_{RD} [1+(\eta - \eta_{RD})] .
\ee
Near radiation/matter equality the scale satisfies
\be
{t \over t_{eq}} =
{2 + \left ( {R \over R_{eq}} - 2 \right ) 
\left ( {R \over R_{eq}} + 1 \right ) ^{1 \over 2}
\over 2 - \sqrt{2}} .
\ee
During matter domination, the density varies inversely
with the spatial volume, so the scale satisfies
\be
R^{3/2}(t) = R_{MD}^{3/2} + 2 R_{MD}^{1/2} c (t-t_{MD}) \quad , \quad
R(\eta) = R_{MD} \left [ 1+{1 \over 2}(\eta - \eta_{MD})\right ]^2 .
\ee

The proper distance to the particle horizon satisfies
\be
d_H = \int _0 ^{d_H} \sqrt{g_{rr}} dr =
R(t) \int_0 ^t {c dt' \over R(t')} + d_H (t_1) .
\ee
During inflation this has the form
\be
d_H = {c \over H_I} \left ( e^{H_I (t-t_I)} -1
  \right ) + d_I ,
\ee
during radiation domination it satisfies
\be
d_H(t) = R(t) \left (
{R(t) - R_{RD} \over R_{RD}}
\right ) + d_H(t_{RD})
\Rightarrow 2 c (t-t_{RD}) + d_{RD} ,
\label{radhorizon}
\ee
and during matter domination
\be
d_H(t) = 2 R(t) \left (
{R^{1/2}(t) - R_{MD}^{1/2} \over R_{MD}^{1/2}}
\right ) + d_H(t_{MD})
\Rightarrow 3 c (t-t_{MD}) + d_{MD},
\ee
where the arrows demonstrate the functional
behaviors well within the given epoch.

If $R$ is an arbitrary scale in the
Friedmann-Lemaitre equations for a spatially flat space in the
radiation-dominated epoch and a phase transition occurs at scale
$R_{PT}$ with expansion rate ${\dot R}_{PT}$,
then Eq. \ref{Hubble_eqn} gives
\be
{\dot R}R = {\dot R}_{PT}R_{PT}=const.
\ee
This form can be integrated to give typical time scales of the form
$\Delta t = {R \over2 {\dot R} }$.
Any \emph{quantized} energy scale E defines a
length scale $R_{\epsilon}$ by the relation
\be
E = {\hbar c \over R_E}. 
\ee
Using a quantized energy
scale  of order E and cosmological time associated with this
scale  in the
energy-time uncertainty relation defines a constraint on
the expansion rate associated with that scale:
\be
\Delta E \Delta t \geq {\hbar \over 2}  \Rightarrow 
{\hbar c\over R_E} {R_E \over 2 \dot{R}_E} \geq {\hbar \over 2}
\Rightarrow \dot{R_E} \le c.
\ee 

We can utilize the scale invariance of the FL equations to
examine the subsequent evolution of scales $\tilde{R}$
that progressively satisfy the measurability constraint
$\dot{\tilde{R}}=c$.  During the radiation epoch, the density
varies with the inverse square of the time $t$:
\be
\left ( {\dot{\tilde{R}} \over \tilde{R}   }   \right ) ^2 =
{8 \pi G_N \over 3 c^2} \rho_{UV}
\left ( { t_{UV} \over t  }   \right ) ^2 =
\left ( {c \over R_{UV} }   \right ) ^2 \left ( { t_{UV} \over t  }   \right ) ^2 .
\ee
Thus the scales and modes satisfy the measurability
condition $\dot{\tilde{R}}=c$ at times
\be
\tilde{R} = R_{UV} \left ( {\tilde{t} \over t_{UV}} \right )
\quad , \quad
\tilde{k} = k_{UV} \left ( {  t_{UV} \over \tilde{t} } \right ) .
\ee
Substitution for $t_{UV}=R_{UV} / 2c$ demonstrates that the threshold
for modes that satisfy the measurability condition follow
the evolution of the horizon in Eq. \ref{radhorizon}. 

Therefore, assuming that the scale factor at the time
of the phase transition is defined by a quantized microscopic
energy scale $R_{PT}=R_E$,
this scale must satisfy\cite{SLACTalk}
\be {\dot R}_{PT} \: = \: c . \ee
Setting the expansion rate to $c$ in the Lemaitre equation \ref{Hubble_eqn}
with $\kappa=0$,
the energy density during dark energy UV de-coherence is given by
\be
\rho_{UV} \: = \:
{3 c^2 \over 8 \pi G_N}  \left ( {c  \over R_{UV} } \right )^2 \, - \, \rho_\Lambda.
\label{rhoFL}
\ee
For convenience, energy scales $m_{UV}$ and $\epsilon$ will be defined
from these densities using
\be
\rho_{UV} \equiv {(m_{UV} c^2) ^4 \over (\hbar c)^3} \quad \quad ,
\quad \quad \rho_\Lambda \equiv {\epsilon  ^4 \over (\hbar c)^3} \quad .
\ee
Similarly, the scale acceleration at the time of this transition can
be determined:
\be
{ \ddot{R}_{UV} \over R_{PT} } \: = \:
- c^2 \left ( {1 \over R_{UV} ^2 }+ {\Lambda \over 3}
\right )
\Rightarrow  \ddot{R}_{UV} \cong -{c^2 \over R_{UV}}.
\label{PTaccel}
\ee

\subsection{Spatial Curvature Constraints}
\indent

The energy density during UV de-coherence $\rho_{UV}$ can be directly determined
from the Lemaitre equation \ref{Hubble_eqn} to satisfy
\be
H_{UV} ^2 =
\left( {c \over R_{UV}} \right )^2 \: = \: {8 \pi G_N \over 3 c^2} 
\left ( \rho_{UV} + \rho_\Lambda \right ) -
{\kappa c^2 \over R_{UV} ^2}.
\label{decoherence}
\ee
A so called
``open" universe ($\kappa=-1$) is excluded from undergoing this transition, since
the positive dark energy density term $\rho_\Lambda$ already excludes a solution
with $\dot{R} \leq c$.
Likewise, for a ``closed" universe that is initially radiation dominated,
the scale factors corresponding to de-coherence $\dot{R}_{UV}=c$
and maximal expansion $\dot{R}_{max}=0$ can be directly compared.
From the Lemaitre equation
\be
{c^2 \over R_{max}^2} \: = \: { 8 \pi G_N \over 3 c^2} \left [ \rho (R_{max}) + \rho_\Lambda \right ] \: \cong \:
{ 8 \pi G_N \over 3 c^2} \rho_{PT} {R_{PT} ^4 \over R_{max} ^4} \Rightarrow
R_{max}^2 \: \cong \: 2 R_{PT} ^2 .
\ee
Clearly, this closed system never expands much beyond the transition scale.
Quite generally, the constraints consistent with quantum measurability for quantized
energy scales ($\dot{R}_{UV} \le c$)
\emph{requires} that all cosmologies which develop structure be spatially flat.

The evolution of the cosmology during the period
for which the dark energy UV scale is in the
microscopic thermal spectrum is
expected to be accurately modeled using the FL equations.  There is a period of
deceleration, followed by acceleration towards an approximately De Sitter expansion.
The rate of scale parameter expansion is sub-luminal during a
finite period of this evolution, as shown in Figure \ref{redsrate3}.
\onefigure{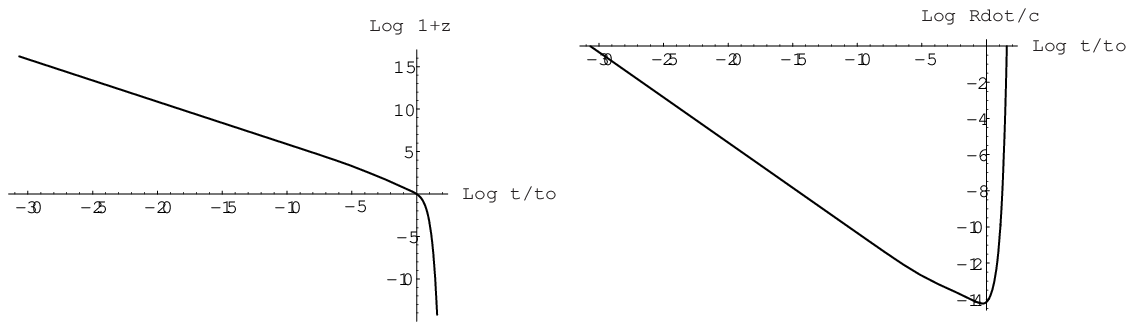}{Log graphs of redshift $R_o/R$ and gravitational UV scale expansion rate vs time}
The particular value for the scale at de-coherence (which is determined by
the microscopic dynamics of the dark energy during de-coherence)
chosen for the graphs
is given in terms of the measured dark energy density
$\rho_\Lambda = \epsilon^4$.
The present time since the ``beginning" of the expansion
corresponds to the origin on both graphs.
The value of the expansion rate is by assumption equal to the speed of light for
any particular value chosen for $R_{PT}$, as well as when this expansion
scale reaches the de Sitter radius $R_\Lambda\approx 1.56 \times 10^{28} cm$
associated with the measured dark energy.

\subsection{Gravitating quantum energy scales}
\indent

The existence of a cosmological bound on the density
attainable by non-coherent, thermal energies has
implications on limits of applicability of classical
relativity.  This suggests that fundamental microscopic physics
undergoes a macroscopic quantum transition for densities
beyond this limit, modifying the assumptions of classical
relativity in this regime.  It is well known that coherent particles
violate the predictions of classical relativity (just as
the coherent physics of stationary atoms violate the predictions
of classical electromagnetism).  For instance, the horizon
of a black hole with mass $M$, spin $S \equiv J M c$, and
charge $q^2 \equiv Q^2 c^4 / G_N$ is given by
\be
r_H = R_S \pm \sqrt{R_S ^2 - (J^2 + Q^2)}  \quad , \quad
R_S = {2 G_N M \over c^2} .
\ee
The term under the radical is unphysical for an electron,
\be
R_S ^2 - (J^2 + Q^2) = 
4 \left( {m_e \over M_P} \right ) ^2 L_P ^2 -
\left [  \left ( {\hbar \over 2 m_e c}  \right ) ^2  +
\alpha L_P ^2  \right ] \cong 
- \left ( {\hbar \over 2 m_e c}  \right ) ^2 <0 ,
\ee
being dominated by the quantized angular momentum.  Quantum
behavior therefore modifies classical general relativity in the
regions near the horizon.

A useful estimate can be obtained with regards to the expected
mass and distance scales separating coherent from non-coherent
behavior.  Writing $M_{CL} c^2 \equiv {4 \over 3} \pi R^3 \rho_{UV}$
as the mass scale representing this
coherence limit in a Schwarzschild geometry, and introducing
a microscopic critical mass scale $\rho_{UV} \equiv
{(m_{UV} c^2)^4 \over (\hbar c)^3}$, an upper limit for a
coherence radial scale is given by
\be  
R_{CL} = \left ( {3 \over 4 \pi}   {M_{CL} \over m_{UV}} \right ) ^{1/3}
\lambda_{m_{UV}}.
\ee
Later, we will estimate the UV energy scale as
$m_{UV} c^2 \cong 3 \times 10^3 GeV$.  Comparing this
radial scale with the Schwarzschild radius for a classical black
hole, coherence transition scales $R_{CL*}, M_{CL*}$ can be calculated:
\be
\begin{array}{l}
R_{CL} \cong  \left ( {M_{CL} \over M_{Sun}}  \right ) ^{1/3}
\times 42 cm , \\
R_{CL*}\cong 3 \times 10^{-8} R_{S,Sun} \cong 9 \times 10^{-3} cm , \\
M_{CL*}\cong 3 \times 10^{-8} M_{Sun} \cong 3 \times 10^{30} M_P .
\end{array} 
\ee
Thus, if the early cosmology is any indication, thermal energies
with Schwarzschild radii less than about 0.01 cm should show
macroscopic coherence effects.

For quantum thermal systems, typical thermal energies
$k_B T_{crit}$ are given by kinetic energies for constituent
particles of mass m, which
define a thermal distance scale $R_{thermal} \approx
{\hbar c \over k_B T_{crit}}$ that satisfies
\be
R_{thermal} \lambda_m \sim (\Delta x)^2
\label{thermalscale}
\ee
in terms of the Compton wavelength of the mass scale
$\lambda_m \equiv \hbar / mc $ and
the scale of zero-point motions of those masses.
This relationship just follows from the momentum-space
uncertainty principle.  For example, for a degenerate free
Fermi gas, the number density relationship
$n={g_m \over 6 \pi^2} (2 m \epsilon_{thermal})^{3/2}$
implies $R_\epsilon \lambda_m = {1 \over 2}
\left ( {6 \pi^2 \over g_m} \right )^{2/3} (\Delta x)^2$\cite{standardtext}.
For a simple harmonic oscillator, the zero-point energy
satisfies ${({\Delta p})^2\over 2 m} + 
{1 \over 2} m \omega ^2 (\Delta x)^2={1 \over 2} \hbar \omega$. 
The zero-point kinetic and potential energies each partition
half of the vacuum energy ${1 \over 4} \hbar \omega$,   The resulting uncertainties
$(\Delta x)^2 = \hbar / 2 m \omega$ and $(\Delta p)^2=
\hbar m \omega /2$ saturate the quantum measurability
condition $\Delta x \Delta p = \hbar/2$.  Thus, for simple
harmonic motions $R_{\hbar \omega} \lambda_m = 2 (\Delta x)^2$.
For systems with fixed microscopic coherence scales a condensate
loses macroscopic coherence below a critical density as illustrated
by Fig. \ref{critdens3}.
\onefigure{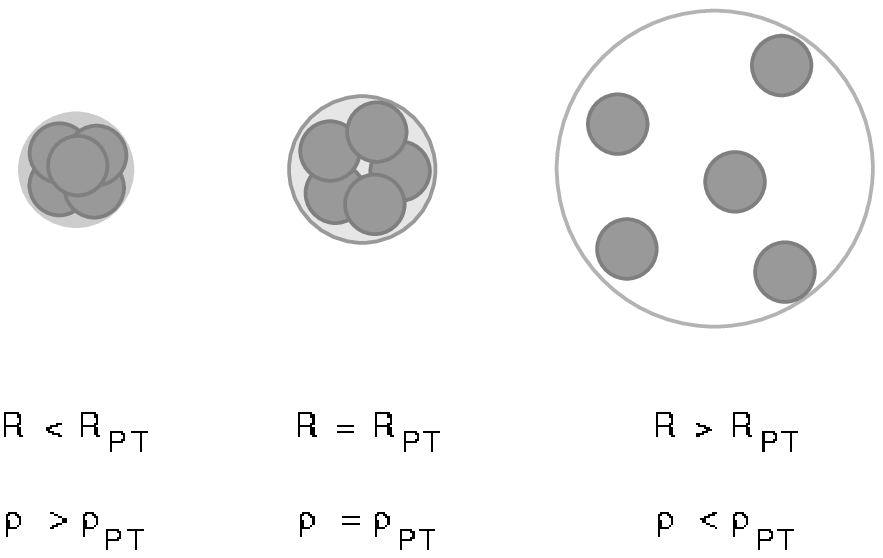}{Overlapping
regions of coherence during expansion} 

\subsection{Estimate of size of source masses for the zero-point energy}
\indent

To estimate the energy scales associated with the zero-point
motions, assume there are
$N$ such sources in a volume specified by $R_{\epsilon} ^3$, and
the energy parameter $\epsilon$. 
On average, each coherent energy unit contributes zero-point
energy of the order
\be
{\epsilon \over N} \sim 
{(\Delta P)^2 \over 2M} \geq {\hbar^2 \over 8M (\Delta X)^2},
\ee
where the uncertainty principle has been used in the form
$(\Delta P)(\Delta X) \geq {\hbar \over 2}$.  Replacing the spatial
uncertainty with the coherence scale $\Delta X \sim R_\epsilon$
relates the energy scale to the zero-point energy $M \sim N \epsilon$.
The cosmological density at the time of the phase transition
is given by the ratio of total (non-relativistic) energy to the volume
of coherence
\be
\rho_{UV} =
{NMc^2\over R_{UV}^3} \sim  N^2 {\epsilon^4 \over
(\hbar c)^3} = N^2 \rho_{\Lambda}.
\ee
The FL equations determine the density during the
phase transition from Eq. \ref{rhoFL} 
\be 
\rho_{UV} \cong  {3\over 8 \pi}{(M_{P}c^2 \epsilon^2)^2 \over
(\hbar c)^3} 
\ee 
giving direct estimates of the coherent energy units
involved in the zero-point motions
\be
N \sim {M_{P}c^2\over \epsilon} \quad ; \quad
M \sim M_{P}.
\ee
That is, the sources of the vacuum
energy must be at the Plank mass scale, each with zero
point energy $\sim {\epsilon^2 \over M_P c^2}$ on average. 
The coherent mass units undergoing zero-point behaviors
have pairwise gravitational couplings $G_N M^2$ of order unity,
whereas the de-coherent energy density during the FL
expansion will consist of masses with considerably smaller
gravitational couplings. Any microscopic mass $m$ with the coherence scale defined
in Eq. \ref{thermalscale} will have zero-point energies of the order
${(\Delta P)^2 \over 2 m}  \sim \epsilon$ at the time of the phase transition,
 which is expected to red shift as the cosmology expands.
If the space-time were Schwarzschild, the scale $R_{UV}$
would be that associated with the Schwarzschild
radius of that geometry.
For brevity, the collective modes of the Planck
mass units will here be referred to as \textit{gravons}.  The zero-point motions
of those coherent energy units correspond to the vacuum energy
of the gravons.

%% file: subfluc.tex
\section{Estimate of Density Fluctuations
\label{PTfluc}}
\indent

The question arises as to why the dark energy
scale freezes out of the subsequent cosmological
expansion.  This scale must be a microscopic (non-expanding) scale
of gravitational relevance.  The energy scale is expected
to freeze out because the medium in some way dissolves
(or more precisely, precipitates)
at the gravitational UV scale.  In this sense 
it is the FL energy density $\rho_{UV}$ that
de-coheres from the dark energy $\rho_\Lambda$
(making our reference of this process as dark energy
de-coherence somewhat a misnomer, since actually
the thermal energy scales de-cohere).  We will
explore scenarios that might decouple
the subsequent expansion from the dark energy
scale.

If the initial cosmology is inflationary, one
expects the Hubble rate at the UV scale to
determine $\rho_{UV}$ (which represents the vacuum energy
density of the thermal state), assuming through
energy conservation that the constant inflationary energy density
thermalizes into the radiation energy density
that drives the subsequent expansion.
The dark energy of the present cosmology is
due to the local thermal effects of the inflationary
deSitter horizon prior to the end of inflation.
If the microscopic density scale associated
with the transition is $\rho_{v}$, then the
fluctuations scale for a transition from
an inflationary to a microscopic
thermal state is expected to be of the
order ${\rho_v \over \rho_{UV}}$.
The horizon is expected to dissolve due to
the onset of the microscopic thermalizations. 
More on this will be discussed in Section \ref{inflationsection}.

A second scenario involves a pure quantum transition
from an initial quasi-stationary state.
As the zero-point motions locally de-cohere,
the deviations from uniformity are expected to appear as
fluctuations in the cosmological energy density. 
Since these motions are inherently
a quantum effect, one expects the fluctuations to exhibit the
space-like correlations consistent with a quantum phenomenon. 
Measurable effects of quantum mechanical de-coherence
are expected to manifest stochastically.
As an intuitive guide into how dark energy might freeze
out as the system de-coheres, consider a uniform distribution
of non-relativistic masses $M$ interacting pairwise through simple harmonic
potentials at zero temperature.  
If charges were used as field sources rather than masses,
Bohr and Rosenfeld\cite{BohrRosenfeld} showed that the
uncertainty relations associated with the positions and momenta
of the charges result in averaged commutation relations between
the electric and magnetic fields which are classically produced by
those charges. 
Correlated zero-point motions of the oscillators
with vanishing time averages
give equal partitioning of energies ${1 \over 4} \hbar \omega$ to the
kinetic and potential components.  If the masses evaporate, the
kinetic component is expected to drive density fluctuations of
an expanding gas, whereas
the potential components remain in the springs (as zero point
tensions) of fixed density, as illustrated in Figure \ref{DCall3}.
\onefigure{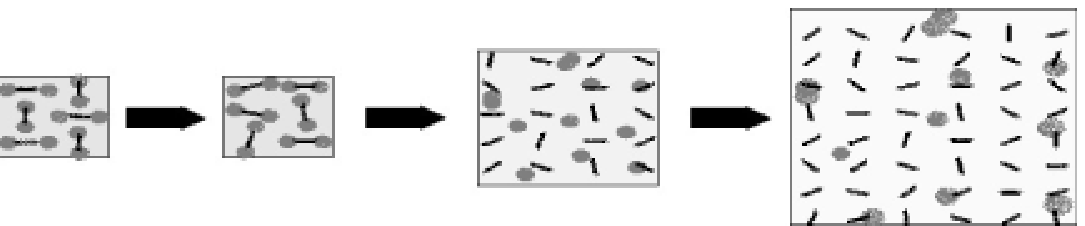}{Early quantum stage undergoing zero-point
fluctuations during de-coherence, with evaporation and
subsequent expansion of thermal energy density associated
with masses which were previously coherently attached
via the springs.  Potential
energy density gets frozen in during evaporation, whereas 
kinetic energy density drives density fluctuations.}
If the evaporation is rapid, the compressional vacuum energy is expected to be frozen in as
dark energy in the background during the phase transition
on scales larger than the de-coherence scale. 
The mass units which were undergoing zero-point motions
no longer behave as coherent masses after evaporation,
and only reflect their former state through the density
inhomogeneities generated during the evaporation process,
no longer coupling to the frozen-in potential component of
the zero point energy other than through a fixed macroscopic
background.  Such an interpretation could
represent the source of a cosmological constant as a frozen energy density,
rather than as a geometric attribute.

An interacting sea of the quantum fluctuations due to
zero point motions should exhibit local statistical variations in
the energy.  Statistical arguments can be made\cite{JLHPNEJ, Poster}
which infer that
the dark energy $E_\Lambda$ is expected to have uniform density,
and to drive fluctuations of the order
\be
<(\delta E)^2> \: = \: E_\Lambda ^2 {d \over dE_\Lambda} <E>
\ee
Given an equation of state $<E> \sim (E_\Lambda )^b$,
the expected fluctuations satisfy
\be
{<(\delta E)^2> \over <E>^2} 
\: = \: b \, {E_\Lambda \over <E>} \, .
\label{delE}
\ee
The specific equation of state depends on the details
of the macroscopic quantum system.  For a gravon gas with
energy $(N_g+ {1 \over 2}) \hbar k_g c=(2 N_g + 1)\epsilon
\sim \rho_{UV} V_\epsilon$,
the exponent in Eq. \ref{delE} has the value $b=1$.
In terms of the densities, one can directly write
${<(\delta E)^2> \over <E>^2} \: = \: {<(\delta \rho)^2> \over \rho^2}
\: = \: b \,  {\rho_\Lambda \over \rho}$.
At the time of the formation of the fluctuations, this means that the
amplitude $\delta \rho / \rho$
is expected to be of the order 
\be
\delta_{PT} \: = \: \left (  b \,{\rho_\Lambda \over \rho_{PT} } \right ) ^{1/2}  
\: = \: \sqrt{b} {R_{PT} \over R_\Lambda}
\label{DelPT}
\ee 
where $\rho_{PT}$ is the cosmological energy density
at the time of the phase transition that decouples the dark energy,
and $\Lambda = 8 \pi G_N \rho_\Lambda /c^4 =3/R_\Lambda ^2$
is the cosmological constant.

To determine how any fluctuations grow,
consider a spatially flat universe with negligible dark
energy satisfying $H^2 = {8 \pi G_N \over c^2} \rho$.  A
small positive (negative) fluctuation in energy density will close (open)
the universe, resulting in a positive (negative) curvature $\kappa$,
with consistent expansion rate, giving
$H^2 = {8 \pi G_N \over c^2} \rho' - {\kappa c^2 \over R^2}$.
Subtracting these equations, and solving for the dimensionless
density perturbations, these perturbations are found to grow
as follows:
\be
{\delta \rho \over \rho} = {3 c^4 \kappa \over 8 \pi G_N}
{1 \over \rho R^2} \propto \left \{
\begin{array}{l}
R^2 \quad radiation \\
R \quad matter/dust
\end{array}  \right . .
\ee
Thus, for adiabatic perturbations (those that fractionally perturb the number
densities of photons and matter equally),
the energy density fluctuations grow according to\cite{PDG}
\be
\delta \: = \: \left \{
\begin{array}{cc}
\delta_{PT}  \left ( {R(t) \over R_{PT}} \right ) ^2  &
radiation-dominated \\
\delta_{eq} \left ( {R(t) \over R_{eq}} \right ) &
matter/dust-dominated
\end{array}
\right . ,
\ee
which gives an estimate for the scale of fluctuations
at last scattering from a radiation epoch
phase transition expressed by
\be
\begin{array}{l}
\delta_{LS} \, = \, { (1+z_{PT}) ^2 \sqrt{b} \over (1+z_{eq}) (1+ z_{LS}) }
\left (  {\rho_\Lambda \over \rho_{PT} } \right ) ^{1/2} \\ \\  \quad \quad \cong
{1 \over 1 + z_{LS}} \sqrt{{b \, \Omega_{\Lambda o}
\over (1-\Omega_{\Lambda o}) (1+z_{eq})}} \cong 2.5 \times 10^{-5} \sqrt{b}, 
\end{array}
\label{delRad}
\ee 
where a spatially flat cosmology has been assumed.
If the transition occurs during dust/plasma domination,
\be
\delta_{LS}  \: \cong \:
\left ( {1+z_{PT} \over 1+z_{LS}}  \right )
\sqrt{{\Omega_{\Lambda o} b \over 
(1-\Omega_{\Lambda o}) (1+z_{PT})^3 
\left (
1+ {1+z_{PT} \over 1+z_{eq}}
\right ) }} \: ,
\label{delDust}
\ee
which varies from $2 \times 10^{-5}$ if the phase transition
occurs at radiation-matter equality, to $4 \times 10^{-5}$.
These estimates are only weakly dependent on the density during
the phase transition $\rho_{PT}$, and is of the order observed for
the fluctuations in the CMB (see \cite{PDG} section 23.2 page 221). 
Fluctuations of a scale larger than $\delta_{PT}$ would be
correspondingly larger a last scattering.

%% file: scenar06.tex
\section{Scenarios Prior to Phase Transition
\label{scenarios}}
\indent

The period of transition from prior coherence to radiation is
typically referred to as reheating.
Prior to reheating, one expects the energy of the universe to be
in the form of coherent modes of the condensate, which have
available degrees of freedom partitioned differently from
the later thermal state.  During reheating,
the energy in the condensate modes precipitates into a multitude of
particle modes, which thermalize.  

The assumption of radiation dominance during the
phase transition corresponds to a thermal temperature of
\be
\eta(T_{PT}) \: (k_B T_{PT})^4 \: \cong  \:
{90 \over 8 \pi^3} (M_P c^2)^2 \left ( {\hbar c \over R_{PT}} \right ) ^2 ,
\ee
where $\eta(T_{PT})$ counts the number of degrees of freedom associated
with particles of mass 
$m c^2 << k_B T_{PT}$, and $M_P=\sqrt{\hbar c/G_N}$ is the Planck mass.  Here
we have used Eq. \ref{rhoFL} and the energy density for relativistic thermal
energy $\rho_{thermal}= \eta(T) {\pi ^2 \over 30} {(k_B T)^4 \over (\hbar c)^3}$.

\subsection{Inflationary prior state \label{inflationsection}}
\indent

If both classical general relativity driven by the
total energy density and quantum mechanics can
be reliably used to describe the cosmology prior to the
time of the decoupling of the dark energy, quantum measurability constraints
on the gravitational interactions (which have couplings of order
unity) suggests a change in the state of the energy density. 
One scenario demands that this energy density
be in the form of vacuum energy (or zero-point energy) with respect to the forms
of matter/energy prevalent in the cosmology shortly after the transition. 
Inflation then ends when the UV gravitational scale crosses
the deSitter horizon associated with the inflation, initiating
microscopic thermalizations.

Assuming a transition that conserves energy,
if the energy density of the present cosmology during the
phase transition is set by the inflationary  energy
density, the deSitter scale of the inflation
$\Lambda_I=3/R_{\Lambda_I} ^2$ is defined
in terms of the dark energy scale of the present cosmology
using $\rho_{\Lambda_I}=\rho_{UV}$, which gives the UV scale:
\be
\rho_{\Lambda_I} = {\Lambda_i c^4 \over 8 \pi G_N}=
{3 c^2  \over 8 \pi G_N} {c^2 \over R_{UV} ^2}
\Rightarrow R_{\Lambda_I} =R_{UV} .
\ee

The time scale for microscopic thermalizations
could be considerably more rapid than the Hubble time
$\tau << {R_{UV} \over c}$.  We will assume the
microscopic transition behavior to be described by a
functional form $F({t \over \tau})$, where
\be
F(\zeta) \Rightarrow \left \{
\begin{array}{l}
0 \quad \zeta \rightarrow 0 \\
1  \quad \zeta \rightarrow \infty 
\end{array} \right .
\ee
so that during the transition from inflation to radiation,
the energy density is expected to be of the form
\be
\rho(t) = \rho_{UV} \left \{ 1+
\left [  \left ( {R_{UV} \over R(t) }   \right ) ^ 4 -1
\right ] F( {t - t_s \over  \tau })
\right \} .
\ee
Here $t_s$ represents the onset of thermalization. 
Substitution into the LeMaitre equation allows estimation
of the temporal behavior of the scale during thermalization.
During the initial onset of thermalization the expansion is
slightly modified from inflation by
\be
R(t) \cong R_{UV} \left [ 1 +
F'(0) {c \over 4 R_{UV} } {t^2 - t_s ^2 \over \tau}
\right ] e ^ {c(t-t_I) / R_{UV}},
\ee
whereas the behavior is dominated by radiation when
$\Delta t > \tau$, as seen in Fig. \ref{inflaRad}. 
The temporal scale is set by both the
Hubble rate $c/R_{UV}$ and the microscopic scale $\tau$. 
The change in the cosmological scale during
thermalization ${\Delta R \over R_{UV}} \approx {c \tau \over R_{UV}}$
depends on the relative time scales of microscopic vs macroscopic
rates. 
\onefigure{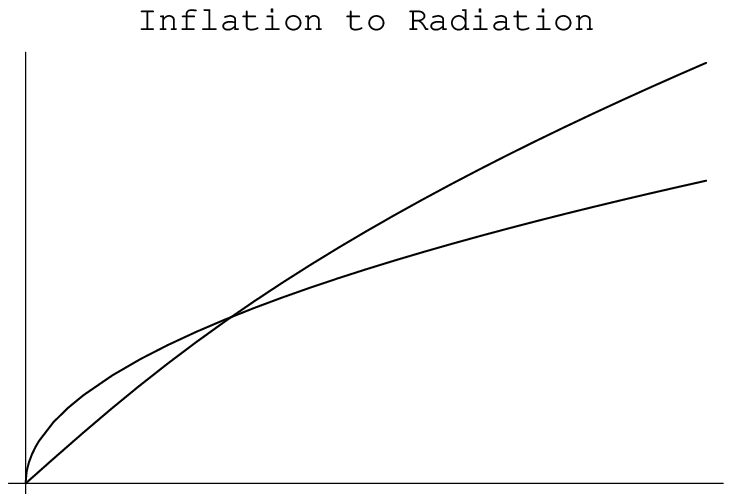}{Inflation to radiation transition vs. pure radiation cosmology. 
Crossover occurs near time $\tau$}.
The figure shows that the initial inflation settles into a
radiation dominated cosmology with a crossover given by $\tau$.

Since the energy density does not red-shift during
the inflationary period, thermalization must represent the
onset of the vacuum modes of the thermal epoch.  This then
implies that the microscopic UV modes $k_{micro}$ satisfy
\be
\rho_{UV} \cong {\eta_{micro} \over 2} {\hbar c \over 8 \pi ^2} k_{micro } ^4,
\ee
where $\eta_{micro}$ represent the degeneracy of microscopic states. 
This implies that in order for the inflation to consistently conserve
energy,
\be
\begin{array}{l}
\hbar k_{micro}c = {\sqrt{4 \pi} \over \eta_{micro} ^{1/4}} m_{UV} c^2
\quad , \quad \textit{microscopic UV scale} \\ \\
\hbar k_{UV}c = {\sqrt{4 \pi} \over g_\epsilon ^{1/4}} \epsilon \quad \quad \quad
\quad \quad , \quad \textit{gravitational UV scale}.
\end{array}
\label{vacuummodes}
\ee
A transition from an inflationary epoch requires the onset of
thermalizations from the microphysics background, implying
the existence of UV scales defined by the transition.

If the energy scale of the microscopic fields is given by 
the electro-weak symmetry breaking scale $v$,
the expected fluctuations would be of the order
\be
\Delta_I \cong \left ( {v \over m_{UV} c^2} \right ) ^4 
\cong 4.5 \times 10^{-5}.
\label{inflafluc}
\ee
This means that if the Higgs field defines microscopic
energy scales, the super-horizon modes should have
amplitudes of the order observed in the CMB.

The transition to a radiation dominated cosmology likely involves a
``latent heat" at the phase transition due to a change
in the entropy per constituent quantum.
If the cosmology transitions from an inflationary epoch,
we can directly estimate the ratio of local thermal entropy to
inflationary entropy.  The entropy during inflation counts
associated super-horizon states (information
lost across the horizon) for given sub-horizon configurations,
and is given by
\be
S_{inflation}={A \over 4 G_N} {k_B c^3 \over \hbar}
=k_B \pi \left ( {R_{UV} M_P c^2 \over \hbar c}  \right ) ^2 ,
\ee
while the deSitter temperature during the inflation is given by
\be
k_B T_{UV}= {\hbar c \over 2 \pi R_{UV}} \sim E_{dark}
\approx  {\pi^2 \over 12 } \hbar k_{UV} v_p.
\ee
For radiation, the entropy density is related to the energy
density by $\sigma_{UV}={4 \over 3} \rho_{UV}$.  Using
the identification for $\rho_{UV}$ given in Eq. \ref{decoherence},
the relative (sub-horizon) entropies during the transition is given by
\be
{S_{thermal} \over S_{inflation}} = {2 \over 3} {\hbar c \over R_{UV}}
{1 \over k_B T_{UV}}
\sim {\epsilon \over m_{UV} c^2} << 1.
\ee
Thus, in this scenario the thermal entropy of the expanding energy density is
smaller compared to that associated with the deSitter horizon of
the prior inflation, although the temperature increases considerably.
The information lost from the existence of an inflationary horizon  is
considerably larger than that lost due to thermalization of microscopic
degrees of freedom.

The scale of the horizon temperature
during the inflation is comparable to the scale of the dark
energy today.  The deSitter thermal energy is due
to super-horizon quantum correlations. 
It is plausible that this thermal energy
which has its roots in quantum correlations
across the horizon
would be connected with the subsequent dark energy.

Using the FL equation for the acceleration Eq. \ref{PTaccel},
the acceleration just after the phase transition is given by
$\ddot{R}_\epsilon= - {c \over \hbar} \epsilon$.  The inflationary scale
just prior to the phase transition has an acceleration
given by $\ddot{R}_\epsilon = +{c \over \hbar} \epsilon$.  The transition
requires a change in the scale acceleration rate of the
order of the dark energy in each scale region of the
subsequent decelerating cosmology.

Others have used related arguments to examine the
initial inflationary period. 
Choosing electro-weak symmetry restoration estimates
of the early 1990's, Ed Jones\cite{Jones90s,Jones97}
predicted a cosmological constant with $\Omega_\Lambda
\simeq 0.6$ before the idea of a non-vanishing small
cosmological constant was fashionable. 
This is one of the appeals of this particular scenario.

\subsection{Quantum Transition cosmology}
\indent

An alternative scenario connects the temporal progression
in classical general relativity to gravitational de-coherence,
implying a macroscopic coherent
quasi-stationary quantum state of density $\rho_{UV}$ prior
to the phase transition.  Since a stationary state is temporally extended,
a stationary density is not expected to drive
the cosmological dynamics in the FL equations
during this ``pre-coherent" period. One cannot
ascertain temporal relations unless the
proliferous time-like interrelations needed to
construct the measurement space-time grid have occurred.
Temporal progression begins only when de-coherent
interrelations break the stationarity of the quantum state,
and is described by the Friedman-LeMaitre equations with
appropriate scale eventually taking the value $R_{UV}$
as the transition proceeds.  
The measurement of time therefore begins with
quantum mechanics, not classical general relativity.
The macroscopic scale evolution is then due to thermal proliferation.
The microscopic gravitational scale gets frozen when vacuum
gravitational energy de-coheres from thermal energies.
An initial quantum stationary state would transition from the (low)
entropy associated with the degrees of freedom of the condensate
to $\sigma_{UV} = {4 \over 3} \rho_{UV}$ during the thermalization transition.

In this description, there is
no initial temporal singularity, or well defined t=0
due to quantum uncertainty.  A
quantum fluctuation of a size that produces dark energy $\epsilon$
with scale $R_\epsilon$ gives the same field dynamics as
would an inflationary scenario with deSitter scale $R_\epsilon$
that matches the phase transition into the decelerating epoch.

%% file: rhopt2.tex
\section{What is Special About $\rho_{PT}$ in \emph{This} Cosmology
\label{specialrho}}
\indent

The currently accepted values\cite{PDG}
for the cosmological parameters involving
dark energy and matter will be used
for the reverse time extrapolation from the present:
\begin{equation}
h_0 \cong  0.72; \ \ \Omega_{\Lambda} \cong 0.73; \ \ \Omega_M \cong 0.27 .
\end{equation}
Here $h_0$ is the normalized Hubble parameter. Note that this
value implies that the universe currently has the critical energy
density
$\rho_c \simeq 5.5 \times 10^{-4} GeV \ cm^{-3}$.
The values
of parameters for the observed cosmology are given by
$\rho_\Lambda \simeq 4.0 \times 10^{-6} GeV/cm^3, \, 
R_\Lambda \simeq 1.5 \times 10^{28} cm \simeq 1.6 \times 10^{10} ly, \,
\epsilon \simeq 2.4 \times 10^{-12} GeV \cong 
\hbar c / 8.4 \times 10^{-3} cm$.

Using Eq. \ref{darkenergy} connecting the dark energy to the UV mode 
$\rho_\Lambda \cong {g_\epsilon \over  b_\Lambda}
{\hbar v_p \over 8 \pi^2} k_{UV} ^4$, 
along with the flatness/measurability requirement
\be
\left ( {c \over R_{UV}}  \right ) ^2 \cong 
{8 \pi G_N \over 2 c^2} \rho_{UV} ,
\label{flatness}
\ee
one only needs to connect the UV mode scale $k_{UV}$
to the FRW scale parameter $R_{UV}$.  The horizon scale mode is
expected to satisfy $|\vec{k}|=\pi / d_H$.  For generality, we will
relate the horizon to the UV scale using
\be
d_{H,UV} = b_R R_{UV} .
\ee
Using Eqns. \ref{darkenergy} and \ref{flatness}, numerical values
can be given for the UV scale parameters:
\be
\begin{array}{l}
d_{H,UV}=b_R R_{UV} \cong  \left (
{g_\epsilon \over b_\Lambda} {v_p \over c}
\right ) ^{1/4} 9 \times 10^{-3} cm \\ \\
\hbar k_{UV} c \cong \left (
{b_\Lambda \over g_\epsilon} {c \over v_p}
\right ) ^{1/4} 6.9 \times 10^{-12} GeV \\ \\
\rho_{UV} \cong b_R ^2 \left (
{b_\Lambda \over g_\epsilon} {c \over v_p}
\right ) ^{1/2} {(3000 GeV)^4 \over (\hbar c)^3} \\ \\
m_{UV} c^2 \cong  b_R ^{1/2} \left (
{b_\Lambda \over g_\epsilon} {c \over v_p}
\right ) ^{1/8} 3000 GeV.
\end{array}
\label{UVparameters}
\ee
The microscopic scale for gravitational de-coherence is
fixed by the critical density $\rho_{UV}$.  If this density
corresponds to the vacuum mode for microscopic states,
the microscopic ultraviolet cutoff can be directly calculated
as in Eq. \ref{vacuummodes}
\be
\begin{array}{l}
\hbar k_{micro}c = \sqrt{4 \pi} \eta_{micro} ^{-1/4} m_{UV} c^2
\quad \quad , \quad \textit{particle modes}, \\ \\
\hbar k_{UV}c = \sqrt{4 \pi} \left ({c  \over v_p g_\epsilon} 
\right ) ^{1/4} \epsilon \quad \quad 
\quad \quad , \quad \textit{gravitational modes},
\end{array}
\ee
where again $v_p$ is the phase velocity of the vacuum modes, and
the particle modes are assumed to have phase velocity $c$.  This gives
an estimate of the microscopic cutoff:
\be
\hbar k_{micro} c \cong  b_R ^{1/2} \left (
{b_\Lambda \over g_\epsilon} {c \over v_p}
\right ) ^{1/8}  \eta_{micro} ^{-1/4} \: 10600 GeV,
\ee
where $\eta_{micro}$ is expected to be of the order ${427 \over 4}$.
The UV scale temperature of the radiation can likewise
be calculated:
\be
\rho_{UV} \sim \eta(T_{UV}) {\pi^2 \over 30} {(k_B T_{UV})^4 \over
(\hbar c)^3} \Rightarrow  
k_B T_{PT} \sim b_R ^{1/2} \left (
{b_\Lambda \over g_\epsilon} {c \over v_p}
\right ) ^{1/8} 1250 GeV  ,
\ee
which corresponds to a redshift $z_{UV} \sim 1 \times 10^{16}$.

In most of our calculations we will assume scalar modes $g_\epsilon =1$.
For a transition from inflation, the horizon scale inherently satisfies
$\dot{R}_I=c$, and is expected to directly correspond to the UV scale,
setting the parameter $b_R=1$, and all vacuum energy will be assumed
to constitute the dark energy $b_\Lambda=2$.  For the quantum
thermalization scenario, the causal region is taken to satisfy periodic
boundary conditions $b_R = 1/2$, and the compressional vacuum energy
is taken to constitute the dark energy $b_\Lambda=4$.

One expects
microscopic physics to fix the particular scale as a quantum
phase transition associated with the UV energy scale for the
gravitational modes, $\rho_{PT} \equiv (m_{UV}c^2)^4/(\hbar c)^3$.  
As a further illustration of the expectation of the manifestation of
macroscopic quantum effects on a cosmological scale, consider
Bose condensation.  As mentioned in the introduction, since
Compton scattering conserves photon number, the photons in
the early universe should have a non-vanishing chemical
potential, making even these massless particles susceptible to
Bose condensation.
A microscopic critical density is reached when thermal modes can no
longer accommodate a distribution of all of the particles, forcing
macroscopic occupation of the lowest energy state. 
For low mass bosons, the Planck distribution gives the thermal
component of those particles, with macroscopic occupation of
the zero mode for particles that are not thermally accommodated.
For non-relativistic particles of mass $m$, this density satisfies\cite{standardtext}
\be
{N_m \over V} = {\zeta(3/2) \Gamma(3/2) \over (2 \pi)^2 \hbar^3}
(2 m k_B T_{crit})^{3/2} \quad , \quad
\rho_m \cong {N_m \over V} m c^2 .
\ee
For instance, if the scale were associated with the
density of thermal bosonic matter, the critical temperature
is related to the (non-relativistic) mass by
$k_B T_{crit}= \left ( {(2 \pi)^2 \over g_m \zeta(3/2)
\Gamma(3/2)}  \right ) ^{2/3}
m_{UV} c^2 \cong  {6.6 \over g_m ^{2/3}} m_{UV} c^2
\cong 3.313 {(\hbar c)^2 \over (m c^2)^{5/3}}
\rho^{2/3}$.
Since this temperature is comparable to the
ambient temperature of thermal standard model matter at
this density, bosonic matter at this density would have a
significant condensate component.  
Bosons with mass $mc^2 > 2700 GeV$ and densities comparable to $\rho_{UV}$
would be expected to have a condensate component..
Using the relation connecting microscopic and macroscopic scale
given previously by $R_\epsilon \lambda_m \sim (\Delta x)^2$,
the zero-point energies of each of the UV energy units 
$m_{UV}$ is of the order of the dark energy 
${(\Delta P)^2 \over 2 m_{UV} } \sim \epsilon$. 

\subsection{A connection to microscopic physics}
\indent 

Motivated by this
approach, examine symmetry breaking in the early universe \cite{SLACTalk,
Bernstein} with density scale $\rho_{UV}$:
\be
\begin{array}{r}
\mathcal{L} = \sqrt{-g} \left [
-{1 \over 2} (D_\mu \Phi_b) g^{\mu \nu} (D_\nu \Phi_b)
+{1 \over 4} (m_\Phi ^2  + 2 \mu^2) \Phi_b ^2 - {1 \over 8} f^2 \Phi_b ^4
- \rho_{UV} 
\right ]  
 \\  + \mathcal{L}_{particle},
\end{array}
\label{Lagrangian}
\ee
where for an FRW cosmology $\sqrt{-g}=R^3$,
the particle Lagrangian includes the gauge boson field
strength contribution $-F^{\mu \nu} F_{\mu \nu}/ 16 \pi$ ,
and any vacuum energy subtractions are appropriately included in $\rho_{UV}$.
In late times, as usual, the classical solution where the gauge fields
vanish results in a non-vanishing vacuum expectation value
for one of the field components
\be
|<\Phi_1>|=0 \quad , \quad
|<\Phi_2>| \equiv v_\mu={ \sqrt{m_\Phi ^2 + 2 \mu ^2} \over f} .
\label{Higgsvac}
\ee

The energy scale $v$ is expected to be the microscopic
energy scale in late times, when thermal energy states have
cooled to the point that the standard particles have vanishing
ground state expectation values.  The free energy density
$\mathcal{L}$ contains the term $\rho_{UV}$ which is the
initial energy density that thermalizes into the observed
cosmological content. 
The strategy is to use the energy density due to the
symmetry breaking field (Higgs\cite{Higgs}) prior to the
thermalization of its excitations relative to the
background given in Eq. \ref{Higgsvac} and the vector bosons into the
microscopic particulate states whose remnants
persist today.  The action corresponding to this
Lagrangian
\be
W_{matter} \: \equiv \: \int \mathcal{L} d^4 x
\ee
generate the conserved energy-momentum tensor in
the Einstein equation
\be
T_{\mu \nu} \: = \: -{2 \over \sqrt{-g}}
{\delta W_{matter} \over \delta g_{\mu \nu}}.
\label{GRTmunu}
\ee

There is every indication that the cosmology will have
extreme spatial homogeneity during the phase
transition, so that for the present, spatial gradients
will be neglected.  For an FRW cosmology, the Jacobian
factor can be calculated using $g=g_{00}g_{xx}g_{yy}g_{zz}$. 
Using Eq. \ref{GRTmunu} and $\delta \sqrt{-g} = {1 \over 2}
\sqrt{-g} g^{\mu \nu} \, \delta g_{\mu \nu}$, 
the energy density can then be calculated as
\be
T_{00} \: = \: {1 \over 2} ( \dot{\Phi} +  B_0 \Phi )^2 -
{1 \over 4} ( m_\Phi ^2 + 2 \mu^2 ) \Phi_b ^2 +
{1 \over 8} f^2 \Phi_b ^4 + 
\rho_{UV} +
T_{00} ^{particle} .
\label{rhoHiggs}
\ee
Note that as previously mentioned, the initial stationary state
has energy density $T_{00}=\rho_{UV}$.
When particulate degrees of freedom are negligible,
the general temporal equation of motion for the field
is given by
\be
{1 \over R^3} {d \over dt} \left ( R^3 [\dot{\Phi} + B_0 \Phi] \right )
- {1 \over 2} (m_\Phi ^2 + 2 \mu^2 - f^2 \Phi^2)\Phi = 0.
\ee

It is interesting to explore the pre-thermal state
of the quantum fields.
Eq. \ref{rhoHiggs} suggests a gravitational energy
scale of the order $\rho_{UV} ^{1/4} =m_{UV} c^2$.
For independent fields, the density matrix formalizes
a basis independent representation of the state of the
system:
\be
\mathbf{D} = \sum_{j=1}^{\eta} w_j
|\phi_j^R> < \phi_j ^R |.
\ee
Here, the fields $|\phi_j ^R>$ have unit norm, $w_j$
represent the weights of the independent states, and $\eta$
is the number of independent states.  For equal
a-priori probabilities, the weights satisfy $w_j = {1 \over \eta}$.
The microscopic fields in Eq. \ref{rhoHiggs} are expected to
have energy scales normalized by the vacuum expectation
value $v$,
\be
{|\phi_j> \over \rho_{UV} ^{1/4}} = \left (
{v \over m_{UV} c^2}  \right ) |\phi_j ^R>.
\ee
This means that the normalized density matrix for the
pre-thermal physical states is given by
\be
\mathbf{D}= \sum_{j=1}^{\eta} \left (
{v  \over m_{UV} c^2}  \right ) ^2  |\phi_j ^R>< \phi_j ^R | ,
\ee
suggesting that with an equipartition of low mass
modes in the pre-thermal cosmology, the degeneracy
is
\be
\eta= \left ( {v  \over m_{UV} c^2}  \right ) ^2 \approx 
{427 \over 4},
\label{degeneracy}
\ee
where the value has been calculated for
a transition from a quantum quasi-stationary initial state,
with vacuum compressional energy constituting the dark energy.

If both classical general relativity and quantum properties
hold in the earliest stages, the FL equations have an 
initial inflationary period with the scale parameter satisfying
\be
\begin{array}{c}
\left ( {\dot{R} \over R} \right ) ^2 =
{8 \pi G_N \over 3} \rho_{tot} , \\
\rho_{tot} \cong  \left [
{(m_\Phi ^2 c^4 + 2 \mu ^2)^2 \over 8 f^2} +
{1 \over 2} (\dot{\Phi} +  B_0 \Phi)^2 +
{1 \over 8} f^2 \Phi^4 -{1 \over 4} (m_\Phi ^2 + 2 \mu^2)
\Phi ^4 +\rho_{particles}.
\right ]
\end{array}
\ee
The dynamical equation relating the time derivatives of 
the component densities can be obtained from energy
conservation ${T^{0 \mu}} _{;\mu} =0= \dot{\rho}_{tot} +
3{\dot{R} \over R} \: \rho_{tot}$.  This equation describes
the detailed thermalization of energy into the particulate
states of present day cosmology,
and gives the dynamical equation for the
early phase transition.  For the present, assuming a
negligible contribution from the gauge boson, the field
evolves prior to particulate fields using
\be
{1 \over R^3} {d \over dt} (R^3 \dot{\Phi}) -
{1 \over 2}(m_\Phi ^2 c^4 + 2 \mu ^2- f^2 \Phi^2) \Phi =0.
\label{earlyphase}
\ee

If there is inflation to the thermal transition, Eq. \ref{earlyphase}
generates a field of the form
\be \Phi = v_{UV} (1-e^{-H_I (t-t_I)}), \ee
where $v_{UV}$ is the UV scale vacuum expectation value of the field.

If the temporal progression defining the dynamics of
the FL equations only begins after there is de-coherent
energy density present, the initial stationary quantum
state only becomes dynamical once the Higgs field
begins to take a non-vanishing vacuum expectation value. 
For a stationary quantum to thermal transition, the time
evolution begins as the stationary state evaporates
\be
\left ( {\dot{R} \over R} \right ) ^2 =
{8 \pi G_N \over 3 c^2} \rho_{thermal} \approx
\left \{
\begin{array}{l}
0 \quad \quad \quad t \rightarrow 0 \\
{1 \over 4 t^2} \quad \quad t \rightarrow t_{UV} .
\end{array}
\right .
\label{HubbleThermal}
\ee
Suppose the thermal density consists of that proportion
of the energy in the broken symmetry state:
\be
\rho_{thermal} = {\rho_{UV} \over v_{UV} ^4}
 {|\Phi|^4 \over (\hbar c)^3} \Rightarrow
{\dot{R} \over R} = \sqrt{{8 \pi \over 3 \hbar ^2}}
{|\Phi|^2 \over M_P c^2} .
\ee
The initial and end stage of thermalization in Eq. \ref{HubbleThermal}
can be determined:
\be
\Phi \approx \left \{
\begin{array}{l}
{ \hbar \dot{\Phi} \over \mu} 
sinh { \mu t \over \hbar}   \quad \quad \quad \quad \quad
t \rightarrow 0 \\ \\
v_{UV} (1 - e^{-3 H_{UV} t }) \quad \quad
t \rightarrow t_{RD} .
\end{array}
\right .
\ee
As can be seen, the late stage thermalization looks like inflation, but
without the same early time behavior.  The general temporal behavior is
demonstrated in Fig. \ref{thermalize}.
\onefigure{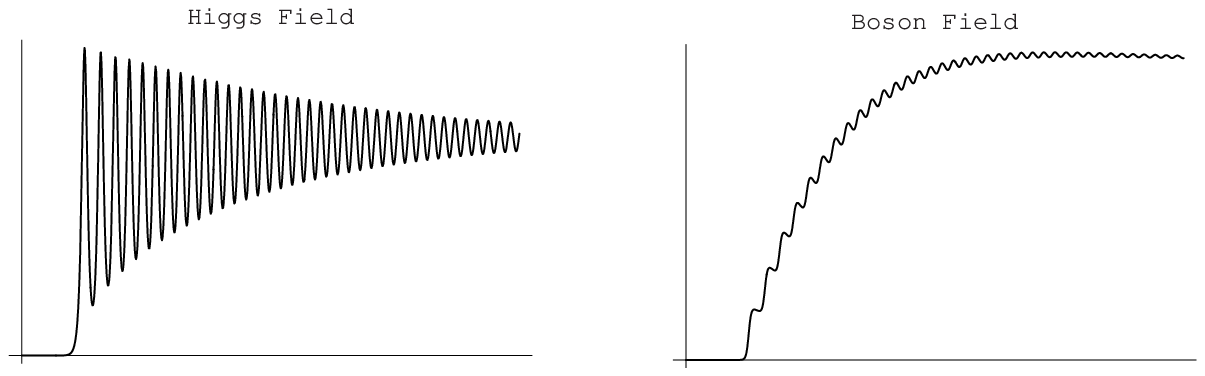}{Higgs field and boson field}
The initial quasi-stationary state eventually undergoes dynamics
on microscopic time scales $\tau \sim {\hbar \over \mu}$, macroscopically
damping to stationary behavior in times related to the cosmological
expansion $H_{UV} ^{-1}$.

We will next briefly examine the thermodynamics of the transition. 
Taking an approach motivated in Reference \cite{Bernstein}, the energy
density in Eq. \ref{rhoHiggs} is taken as
\be
\mathcal{H}=\rho_{UV} + {1 \over 2} 
(\Pi^2 + |\vec{\nabla} \Phi|^2) -
{1 \over 4} (m_\Phi ^2 + 2 \mu^2) \Phi^2 +
{1 \over 8} f^2 \Phi^4 .
\ee
The Noether currents for a global symmetry with
generators $\{ G_r \}$
\be
\sqrt{-g} J_r ^\mu = 
{\partial \mathcal{L} \over \partial D_\mu \psi} G_r \psi
\ee
satisfy ${1 \over \sqrt{-g}} {\partial \over \partial x^ \mu}
\left ( \sqrt{-g} J_r ^\mu  \right ) = 0 = {J_r ^ \mu}_{; \mu}.$
Through co-moving surfaces with vanishing flux of this
current, the charges $\mathcal{N}_r = \int J_r ^0 \sqrt{-g} d^3 x$
are conserved.  The density $\mathcal{H}$ has conserved
charges with generators corresponding to $2 \times 2$ identity
and antisymmetric matrices.  Energy conservation across
the transition requires that only a small constant energy density
$\rho_\Lambda \approx 0$ remain after the symmetry is fully
broken, giving the identification
\be
\rho_{UV} = {1 \over 8} 
{(m_\Phi ^2 c^4 + 2 \mu^2 )^2 \over f^2} .
\ee
The chemical potential is expected to vary through the values
$\mu^4: 0 \rightarrow2 f^2 (m_{UV} c^2)^4
\rightarrow 0$
for the quasi-stationary state through thermalization to the
electro-weak scale.

The excitations satisfy
\be
\begin{array}{l}
E_\pm ^2 = k^2 + 3 \mu^2 + {m_\Phi ^2 \over 2}
\pm \left [ 4 \mu^2 k^2 + \left ( 3 \mu^2 +  {m_\Phi ^2 \over 2}
\right ) ^2 \right ]^{1/2}  \\
\quad \quad \quad \Rightarrow_{k<<\mu} 
\left \{ \begin{array}{c}
\left ( 1 + 
{2 \mu^2 \over 3 \mu^2 + m_\Phi ^2 /2}
\right ) k^2 + 6 \mu^2 + m_\Phi ^2 \quad \quad \textit{massive mode} \\
\left ( 1 - {2 \mu^2 \over 3 \mu^2 + m_\Phi ^2 /2}
\right ) k^2 \quad \quad \quad \quad \quad \quad \quad
\textit{Goldstone mode}.
\end{array}
\right . 
\end{array} 
\ee
and the conserved charge in volume $\Omega$ is given by
\be
\mathcal{N}=- \left . {\partial V \over \partial \mu} \right |_{<\Phi> =v} 
\Omega_{UV} \approx 
{4 \rho_{UV} \Omega_{UV} \over \mu_{UV}},
\ee
which corresponds to the number of particles.
For small $\mu$ (at low density) the charge is given by
$\mathcal{N} \cong \mu v^2 \Omega$, which
implies that at later times
$\mu \approx  \mu_{UV} \left ( {R_{UV} \over R} \right ) ^3$. 
This demonstrates the expected late time vanishing of the
chemical potential as a contribution to the Higgs field
symmetry breaking scale.

We next explore arguments which should connect
the scale of the microscopic fluctuations to macroscopic
parameters.  The fluctuation dissipation theorem in
statistical physics connects the dissipation rate to
stochastic fluctuations of the system.  As brief demonstration
examine a linear response from a stationary solution,
with an additional stochastic gaussian variable
driving fluctuations
\be
{\partial \phi \over \partial t} = 
- \Gamma {\delta L \over \delta \phi} + \zeta(\vec{x},t) ,
\label{flucdiss}
\ee
where the stochastic variable satisfies
$<\zeta(\vec{x},t)>=0, 
<\zeta(\vec{x},t)\zeta(\vec{x}',t')>=D \delta^3(\vec{x}-\vec{x}')
\delta(t-t')$.
The stationary, long time solution connects the RMS fluctuation
of the noise function to the dissipation of the order parameter
and the temperature of the statistical bath  $D=2 \Gamma k_B T$. 
The decays of the Higgs field are expected to drive the
microscopic dissipation of the cosmological field.  Ignoring
the mass differences of the weak bosons, the decay of the
Higgs field excitations into weak bosons are of the order 
\be
\Gamma_{H \rightarrow weak \: bosons} \approx 
{e^2 \over \hbar c} {3 \over 128 \pi sin^2 \theta_W}
\left ( {m_H \over m_W} \right )^2 m_H ^2 c^2 \sim
2.5 \times 10^{-4} \left ( {m_H \over m_W} \right )^2 m_H ^2 c^2
\ee
Setting fluctuations of the order of the microscopic vacuum
expectation value $v$, 
if $k_B T_{UV} \approx 2100 GeV$, we expect $\Gamma \approx 14.5 GeV$. 
If this width corresponds to decays to the whole particle spectrum, we expect
$m_H c^2 \approx 330 GeV$, whereas if the decays are just to weak bosons
$m_H c^2 \approx 719 GeV$

\subsection{Power spectrum considerations}
\indent

Finally, we explore the power spectrum of super-horizon scale
fluctuations to examine the expected behavior of spatial modes
as they satisfy measurability criteria.
For a generic quantity $g$, the relation between the spatial modes
$g_k$ and the power spectrum $P_g$ is given by
\be
\begin{array}{c}
g(\vec{x},t) = \int {d^3 k \over (2 \pi)^{3/2}}
g_k (t) e^{\vec{k} \cdot \vec{x}} , \\ \\
P_g (k) = {k^3 \over 2 \pi^2 } |g_k |^2 .
\end{array}
\ee
For massless scalar fields with an action given by
\be
W= {1 \over 2} \int \partial_\mu \psi g^{\mu \nu} \partial_\nu \psi
\sqrt{-g} d^4 x
\ee
the modes of the fields $\psi_k$ satisfy
\be
\ddot{\psi}_k + 3 {\dot{R} \over R} \dot{\psi}_k +
{k^2 \over R^2} \psi_k = 0.
\ee
Expanding the small perturbations using conformal time
$\psi (\mathbf{x}, \eta)$, the dynamics is given by
\be
{d^2 \psi_k(\eta) \over d \eta ^2 }+
2 {R'(\eta) \over R(\eta) } {d \psi_k(\eta) \over
d \eta} + k^2 \psi_k(\eta) =0.
\label{psieta}
\ee

In an inflation, Eq. \ref{psieta} has a solution
\be 
\psi_k = \Delta_I {e^{-i k \eta } \over \sqrt{2 k}}
\left ( {1 \over R(\eta)}  +
{ i {H_I} \over kc} \right ).
\label{psiinflate}
\ee
The second term in the parenthesis is super horizon scale invariant.
As previously mentioned, fluctuations generated by the electro-weak
(microscopic) scale are expected to produce amplitudes of the order
$\Delta_I ^2 \approx \left ( {v \over m_{UV} c^2} \right ) ^2 \sim 10^{-5}$.
Next, scalar field evolution during radiation domination satisfies
\be
\psi_k = \tilde{\Delta}_{RD} {e^{i k \tilde{R}(\eta)/\tilde{R}_{RD}} \over \sqrt{2k}}
\left ( {\tilde{R}_{RD} \over k \tilde{R}(\eta)}    \right ),
\label{psirad}
\ee
where $\tilde{R}=\tilde{R}_{UV} \left [ 1+(\eta - \tilde{\eta}_{RD})  \right]$
independent of the scale of measurability $\tilde{R}_{RD}$.  The
amplitude is scale invariant in wave number, but decreases with
the inverse of the FRW scale factor.
If the amplitude $\tilde{\Delta}_{RD}$ varies as modes satisfy
measurability constraints and is fixed by the scale of
dark energy de-coherence in Eq. \ref{DelPT}, the amplitude of
the independent modes at last scattering grow to be of the same
order in Eq. \ref{delRad}. 
Finally, consider the growth of field fluctuations during matter
domination. 
A solution can be found that satisfies super-horizon (small k) and 
sub-horizon (large k) mode scales, to first order in the conformal time
so that any scale invariant behavior can be extracted for a transition
during matter domination.  This solution takes the form
\be
\psi_k \Rightarrow_{\eta \rightarrow \eta_{MD}}
{\Delta_{MD} \over \sqrt{2k}} \left (
{4i + 4k \over 3k} - 
{4k - 2i \over 3k}{1 \over 1 + (\eta - \eta_{MD})/2} +
e^{-k(\eta-\eta_{MD})}
\right ) .
\label{psimatter}
\ee
The first term in the parenthesis is seen to contain a contribution
to a scale invariant amplitude, precisely of the order measured for
CMB fluctuations predicted by dark energy de-coherence in Eq. \ref{delDust}.
The power spectrum of super-horizon scale fluctuations is dominated
by a scale invariant term which, if driven by dark energy de-coherence
as previously described, is of the size observed in CMB fluctuations
\be
P_\psi = {k^3 \over 2 \pi^2} |\psi_k|^2 \approx
{8 \over 9} \Delta_{MD}^2
\ee
This result is independent of whether there is an early inflationary
stage. It only depends on a finite dark energy density which requires
only the existence of a UV scale.

\subsection{Conspiracy of coincidences}
\indent

To end this discussion, we will briefly reiterate the numerical
coincidences consistent with the UV dark energy scale assumption.
\begin{itemize}
\item The scale of the fluctuations generated in an inflationary
era (Eq. \ref{inflafluc}), radiation era (Eq. \ref{delRad}), and
matter/dust era (Eq. \ref{delDust})
are found to have an amplitude of the order observed in the CMB.
\item The power spectra of the fluctuations generated during
an inflationary era (Eq. \ref{psiinflate}) and matter/dust era
(Eq. \ref{psimatter}) are demonstrably scale invariant for
super-horizon scale modes.
Fluctuations generated during the radiation era (Eq. \ref{psirad}) are generated in a
manner different from the others, but they grow to a flat amplitude at
last scattering.  However, one needs to question how any of the modes grow
during epochs different from the one in which the fluctuations are produced.
\item The decay width (conversion
rate of condensate energy into thermal energy) during reheating
is often argued to be the same as the expansion rate
(see Riotto\cite{Riotto} Eq. 62).  Using Riotto's formula
this gives a decay width essentially equal to the dark energy at a
cosmological temperature corresponding to $T_{UV}$
(see Eq. \ref{UVparameters}).
\item  The number of microscopic degrees of freedom in the pre-thermal
cosmology counted by equipartition of microscopic ground state expectation
values for independent particle fields is of the observed magnitude in
terms of the particle spectrum (Eq. \ref{degeneracy}).
\item The fluctuations generated during the period of microscopic
de-coherence which are related to the dissipation due to cosmological
expansion are coincidental with the vacuum expectation value of the
Higgs field for electro-weak symmetry breaking (Eq. \ref{flucdiss}). 
\end{itemize}

%% file: discuss2.tex
\section{Conclusions and Discussion}
\indent

This presentation has demonstrated evidence that current
cosmological observations can be accounted for
by the hypothesis that at early times there was
a phase transition from a macroscopically coherent
state which produced dark energy and fluctuations consistent with
those observed today. 
We generally expect some form of cosmological quantum coherence at a scale far from 
the Planck scale, implying space-like correlations and phase coherence.
Gravitational de-coherence at the UV scale defined by the dark
energy gives evidence for the existence of a fixed microscopic
scale of cosmological significance.  
Space-like/supraluminal correlations speak more to the quantum physics of the early 
universe than it does to a classical inflation.

The constraints of quantum measurement and flatness
associates with any quantized energy scale
a cosmological scale whose expansion rate is at most luminal.  
Using the measured value for the dark energy,
some form of microscopic manifestation of gravitational
physics is expected for densities on the TeV energy scale.

Since the results presented depend only on the physics of the
transition, several prior scenarios have been explored which
connect appropriately to the phase transition.  In addition,
numerical coincidences consistent with the various scenarios
have been presented.  In particular, an argument can be made that
for the observed dark energy density, the cosmological density at 
de-coherence generates microscopic fluctuations of the order of the electro-weak 
symmetry scale,  macroscopic damping of the order of the dark energy, and 
fluctuations that are fixed at (or grow to be of) the order observed at last scattering.

%% file: NSBP06.bbl
\begin{thebibliography}{99}

\bibitem{Over74}
A.W.Overhauser and R.Colella, Phys.Rev.Lett. {\bf 33}, 1237
(1974).

\bibitem{Over75}
R.Colella, A.W.Overhauser and S.A.Werner, Phys.Rev.Lett. {\bf 34},
1472 (1975).

\bibitem{WMAP}
C.L. Bennett, et.al., Astrophys.J.Supp. 148, 1 (2003) 

\bibitem{TypeIa}
A.G.Riess et al, Astron.J. 116, 1009 (1998);
P. Garnavich et al, Astrophys.J. 509, 74 (1998);
S. Perlmutter et al, Astrophys.J. 517, 565 (1999)

\bibitem{PDG}
Particle Data Group, {\bf Astrophysics and cosmology}, as posted (2004).

\bibitem{SLACTalk}
J. Lindesay, ``Quantum coherence arguments for cosmological scale",
SLAC-WD-050, presented at SLAC Theory Seminar, 4 May 2005.

\bibitem{Paddy1}
T. Padmanabhan, ``Understanding our universe: Current status and
open issues", invited contribution in \textit{100 Years of 
Relativity- Space-time Structure: Einstein and Beyond}, A.
Ashtekar, ed., World Scientific, Singapore (2005) pp175-204.

\bibitem{Paddy2}
T.Roy Choudhury, T.  Padmanabhan , ``Cosmological parameters from supernova observations: A critical comparison of three data sets", Astron.Astrophys., 429: 807, (2005) [astro-ph/0311622].

\bibitem{densityofstates}
R.H. Lambert, ``Density of states of a sphere and cylinder",
Am.J.Phys.36, 417-420 (1968).  More generally see H. Weyl,
Math.Ann. 71, 441 (1912)

\bibitem{Casimir}
H.B.G.Casimir, Proc.K.Ned.Akad.Wet. 51, 793 (1948)

\bibitem{Lifshitz}
E.M. Lifshitz, Soviet Phys. JETP 2, 73 (1956).  
I.D. Dzyaloshinskii, E.M. Lifshitz, and I.P. Pitaevskii, Soviet Phys. Usp. 4, 153 (1961).
I.D. Landau and E.M. Lifshitz, \textit{Electrodynamics of Continuous Media}, pp368-376
(Pergamon, Oxford, 1960)

\bibitem{Boyer}
T.H. Boyer, Phys. Rev. 174, 1764 (1968)

\bibitem{Kleppner}
D. Kleppner, "With apologies to Casimir," Physics Today, 43,10 (Oct 1990) 9-11.

\bibitem{Wheeler}
J.A.Wheeler and R.P.Feynman, Rev.Mod.Phys. 17, 157 (1945); and Rev.Mod.Phys 21, 425 (1949)

\bibitem{LN2}
J.V.Lindesay and H.P.Noyes, ``Non-Perturbative, Unitary
Quantum-Particle Scattering Amplitudes from Three-Particle
Equations", hep-th/0203262, Found. Phys. 34, 1573-1606 (2004).

\bibitem{BohrRosenfeld}
N.Bohr and L. Resenfeld, ``On the question of the measurability of
electromagnetic field quantities", \textbf{Selected Papers of Leon Rosenfeld},
Cohen and Stachel, eds. ,Reidel, Dordrecht (1979) 357-400, translated from
Mat.-fys. Medd. Dan. Vid. Selsk., 12, no. 8 (1933).

\bibitem{JLHPNEJ}
J.V. Lindesay, H.P. Noyes, and E.D. Jones, ``CMB Fluctuation Amplitude
from Dark Energy Partitions", Physics Letters B 633 (2006) 433-435,
astro-ph/0412477 v3, 7 pages.

\bibitem{Poster}
J. Lindesay and H. P. Noyes, ``Estimate of the CMB Fluctuation Amplitude
from Dark Energy Decoherence", Proc. of the Texas at Stanford
22nd Texas Symposium on
Relativistic Astrophysics at Stanford University, Dec. 13-17 (2004);
SLAC-R-752, eConf:C041213, Poster 1304.

\bibitem{standardtext}
To examine a textbook exposition on degenerate systems,
see R. K. Pathria, \textit{Statistical Mechanics},
Butterworth-Heinemann 2nd ed. (1996), ISBN 0750624698

\bibitem{Riotto}
A. Riotto, ``Inflation and the Theory of Cosmological Perturbations",
Lectures from the Summer School on Astroparticle Physics and
Cosmology, Trieste, hep-ph/0210162 (2002)

\bibitem{Jones90s}
E.D.Jones, conversations with HPN starting in the 1990's when
there was no enthusiasm in cosmological circles fo a ``cosmological
constant" other than zero, let alone a positive one.

\bibitem{Jones97}
E.D.Jones, private communication to HPN c. 1997. The result of
Jones' calculation which gave $\Omega_{\Lambda} = 0.6 \pm 0.1$ is
quoted in {\it Aspects II (Proc. ANPA 20)}, K.G.Bowden, ed.
(1999), p. 207, and cited again in\cite{Noyes01}, p. 561.

\bibitem{Higgs}
P.W. Higgs, ``Spontaneous Symmetry Breakdown without
Massless Bosons", Phys. Rev. 145, No.4 (1966) 1156.

\bibitem{Bernstein}
J. Bernstein and S. Dodelson, ``Relativistic Bose Gas",
Phys. Rev. Let. 66, 683-686 (1991)

\end{thebibliography}
